\documentclass[aps,pre,twocolumn,reprint,nofootinbib,nobibnotes,showpacs,groupedaddress]{revtex4-1}
\pdfoutput=1
\usepackage{subfigure}
\usepackage{amsmath, amssymb}
\usepackage{color}
\usepackage{graphics}
\usepackage{epsfig}
\usepackage{natbib}
\usepackage{url}
\usepackage{bm}
\usepackage{verbatim}

\newcommand{\etal}{\textit{et al.}}
\renewcommand{\vec}[1]{\bm{#1}}

\newcommand{\beq}{\begin{equation}}
\newcommand{\eeq}{\end{equation}}
\newcommand{\beqal}{\begin{align}}
\newcommand{\eeqal}{\end{align}}

\begin{document}

\title{Magnetic helicity and the evolution of decaying magnetohydrodynamic turbulence}
\author{Arjun Berera}
\email[]{ab@ph.ed.ac.uk}
\affiliation{SUPA, School of Physics and Astronomy, University of Edinburgh, Mayfield Road, EH9 3JZ, UK}

\author{Moritz Linkmann}
\email[]{m.linkmann@ed.ac.uk}
\affiliation{SUPA, School of Physics and Astronomy, University of Edinburgh, Mayfield Road, EH9 3JZ, UK}

\begin{abstract}
Ensemble averaged high resolution
direct numerical simulations of reverse spectral transfer
are presented, extending on the many single realization numerical
studies done up to now.  This identifies this type of spectral transfer as a statistical
property of magnetohydrodynamic turbulence and thus
permits reliable numerical exploration of
its dynamics.
The magnetic energy decay exponent from these ensemble runs
has been determined to be
$n_E = (0.47 \pm 0.03) + (13.9 \pm 0.8)/R_{\lambda}$ for initially helical 
magnetic fields.
We show for the first time that even after removing the
Lorentz force term in the
momentum equation, thus decoupling it from the
induction equation, reverse spectral transfer still
persists.  The induction equation is now linear with an externally
imposed velocity field, thus amenable to numerous analysis
techniques. A new door has opened for analyzing reverse spectral transfer, with
various ideas discussed.
\\
\\
\noindent In Press Physical Review E Rapid Communication (2014).
\end{abstract}


\maketitle

On large length scales kinetic plasma effects can be neglected and 
magnetohydrodynamics (MHD)
gives a good first order approximation to plasma evolution.   
The relevance of MHD turbulence ranges from industrial application, 
fusion research, solar physics (\emph{e.g.}~coronal heating) to astrophysics
and cosmology, 
where it might leave detectable signatures in astrophysical 
processes \cite{Beresnyak12a}
and even for the very early 
universe the possibility of a large-scale primordial magnetic field 
\cite{Brandenburg96,Son99,Tevzadze12,Kahniashvili12}. 
While there are many applications of MHD turbulence research in the above areas,
some of the theoretical problems still remain open.
Fundamental research in MHD turbulence consists of many active fields such as 
the amplification of a seed magnetic field by dynamo processes 
\cite{Tobias13},
different proposed models concerning the scaling of the energy spectra
taking small-scale anisotropy into account 
\cite{Iroshnikov63},  
and MHD turbulence decay to name only a few.  

Selective decay \cite{Mininni13,Biskamp93},
that is the decay of ideal quadratic invariants of MHD flows 
at different rates, dominates the nonlinear evolution of decaying 
turbulent MHD flows. 
It is related to the direction of spectral transfer of said ideal
invariants.
The magnetic helicity, which is one of the three ideal invariants of MHD 
flows (the other two being the total energy and the cross helicity), has been
shown to influence the evolution of the magnetic field \cite{Biskamp99,Mueller00}  
possibly through its reverse spectral transfer (RST)
\footnote{
The alternative terminology ``inverse
cascade" might be inaccurate as it implies spectral locality \cite{Mueller12}.
} \cite{Pouquet76}.
An understanding of the underlying
mechanism of RST remains elusive, though much progress has been made.

In this paper we expand the numerical study of RST and its effects on MHD 
turbulence decay to ensemble-averaged
data, which permits more reliable numerical exploration of the MHD equations 
compared to the single realization studies done up to now.
We also deconstruct the nonlinear MHD equations and identify some of
the underpinnings of RST. Here we treat incompressible MHD turbulence 
only, and the magnetic Prandtl number is set to unity.
 
The incompressible decaying MHD equations are
\begin{align}
\label{eq:momentum}
\partial_t \vec{u}&= - \frac{1}{\rho}\nabla P -(\vec{u}\cdot \nabla)\vec{u}
 + \frac{1}{\rho}(\nabla \times \vec{b}) \times \vec{b} + \nu \Delta \vec{u}  \ , \\
\label{eq:induction}
\partial_t \vec{b}&= (\vec{b}\cdot \nabla)\vec{u}-(\vec{u}\cdot \nabla)\vec{b} + \eta \Delta \vec{b}\ , \\
\label{eq:incompr}
&\nabla \cdot \vec{u} = 0 \ \ \mbox{and} \ \  \nabla \cdot \vec{b} = 0 \ ,  
\end{align}
where $\vec{u}$ denotes the velocity field, $\vec{b}$ the magnetic induction expressed in Alfv\'{e}n units, 
$\nu$ the kinematic viscosity, $\eta$ the resistivity, $P$ the pressure and $\rho=1$ the density.
Equations \eqref{eq:momentum}-\eqref{eq:incompr} are numerically solved 
in a cubic domain of length $L=2 \pi$ with periodic boundary conditions
using a fully de-aliased pseudospectral MHD code,
which we developed extending the hydrodynamic code of \cite{Yoffe12}. 
All simulations  
satisfy $k_{max}\eta_{mag,kin} \geqslant 1.26$, where 
$\eta_{mag,kin}$ are the Kolmogorov scales associated with the magnetic and 
velocity fields, respectively.
We do not impose a background magnetic field, and both the initial magnetic 
and velocity fields 
are random Gaussian with zero mean, with initial magnetic and kinetic energy 
spectra $E_{mag,kin}(k)\sim k^4\exp(-k^2/(2k_0)^2)$,  
unless otherwise specified. The peak wavenumber $k_0$ is varied for different 
simulations depending on the desired scale separation and resolution 
requirements. 
The initial relative magnetic helicity is $\rho_{mag}(k)=kH_{mag}(k)/2E_{mag}(k)=1$, 
the initial cross helicity is negligible 
and the initial velocity field is non-helical, 
unless 
otherwise specified.
The ratio between magnetic and kinetic energies $\Gamma(t)=E_{kin}(t)/E_{mag}(t)$ 
equals unity at $t=0$, where $E_{mag,kin}(t)=\int dk \  E_{mag,kin}(k,t)$.  
All spectral quantities have been shell- and ensemble-averaged.
Results have been obtained for a range of Reynolds numbers; the figures 
show data from the highest resolved simulations only. 
A summary of simulation details is shown in Table \ref{tbl:simulations}, 
further details including benchmarking against results in the literature 
\cite{Mininni06, Morales12} can be found in \cite{SM}. 

The numerous single realization studies of RST make it evident that it
should also appear as a property in ensemble-averaged data, but this
is the first analysis to adopt this procedure.
In isotropic hydrodynamic turbulence it is pointed out in \cite{Sagaut08} 
that the direct cascade of kinetic energy is 
an ensemble-averaged concept and in \cite{McComb14a} that 
a single realization could show energy transfer towards small wavenumbers, 
and it is the mean kinetic energy transfer that proceeds from low to 
high wavenumbers.
At low $k$, the regime important for RST, shell averaging
is not an optimal averaging method, since
there is only a small number of points to average over. 
Furthermore, the modes in a given $k$-shell do not evolve independently from 
each other, as they become increasingly correlated by nonlinear mode coupling,
whereas different realizations in an ensemble are statistically 
independent. It has also been noted that the actual measuring process in
experimental studies of decaying turbulence results in an ensemble
average \cite{Mininni13}.

\begin{figure}[ht]     
   \includegraphics[width=\columnwidth]{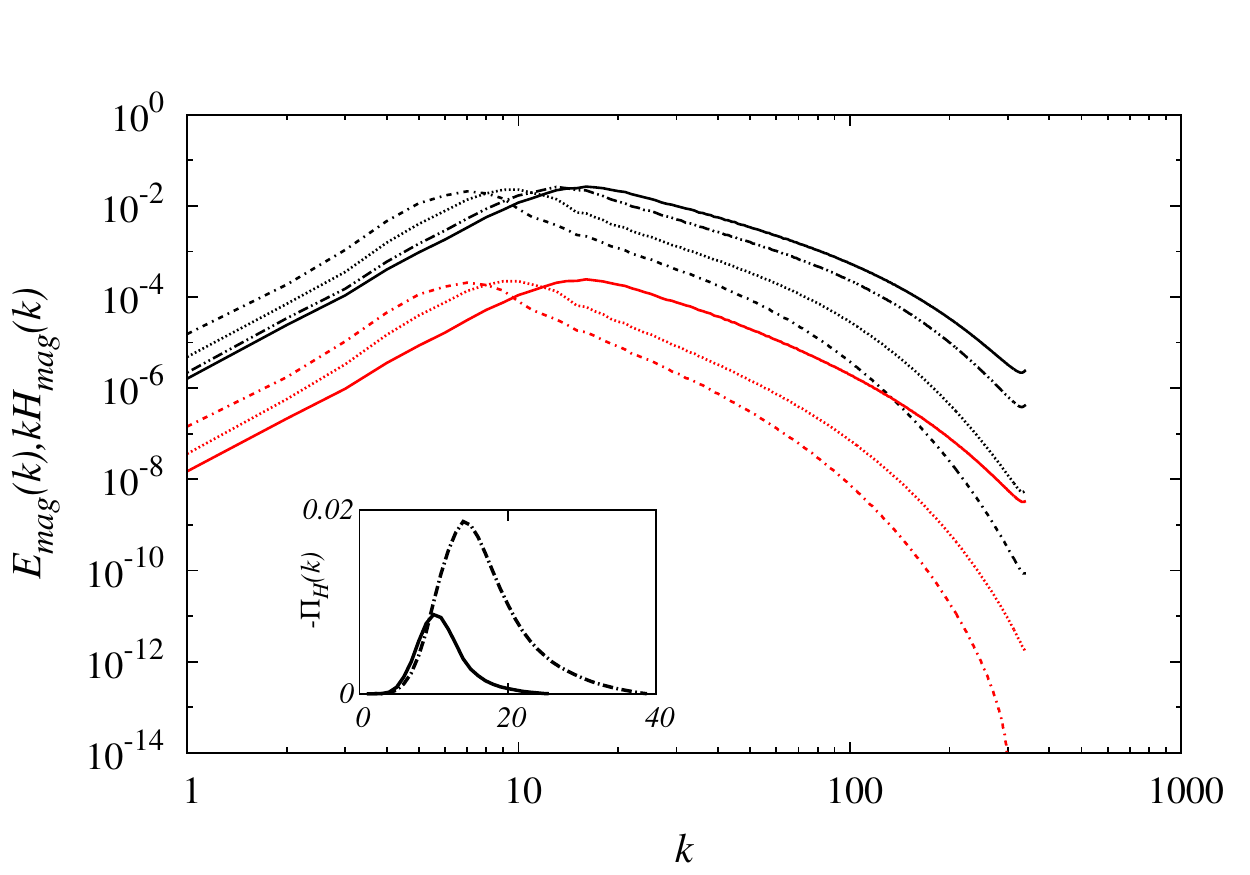} 
\caption{
(Color online) Magnetic energy and helicity spectra of run H9 showing reverse spectral transfer. The 
black (upper) lines refer to $E_{mag}(k)$, red (lower) lines to 
$kH_{mag}(k)$. Solid lines 
indicate one initial turnover time $t_0$, dotted and dash-dotted lines 
refer to $2t_0$, $5t_0$ and $10t_0$. 
The inset shows the flux of magnetic 
helicity at $t_0$ and $5t_0$.
}
\label{fig:ic}
\end{figure}

Our ensemble-averaged results for $E_{mag}(k)$ and $H_{mag}(k)$ at 
different times are shown in
Fig.~\ref{fig:ic} for run H9 in Table \ref{tbl:simulations},
where the helicity spectra have been shifted for easier comparison.
Error bars have been
omitted to facilitate visual comparison between spectra at different times,
but it should be noted that
the measured spectra do not lie within the error of one another. 
The inset of Fig.~\ref{fig:ic} shows the flux of magnetic helicity 
$-\Pi_H(k)= \int_0^k dk' \ T_H(k')$, where $T_H(k')$ denotes the transfer spectrum 
of the magnetic helicity \cite{Biskamp93}, at one and five initial large eddy turnover 
times $t_0$ in the very low $k$
 region. It is positive (as is $H_{mag}(k)$ at all times), which again indicates RST, 
but not constant. 
This indicates the absence of an inertial range, hence the observed 
RST here cannot be named a cascade.  
This is in accord with standard results in this wavenumber range; an inertial 
range is not expected at the very low $k$ \cite{Mueller12}. 
For visual purposes, we show a low $R_{\lambda}$ result, which allows for higher 
scale separation at the low wavenumbers. Higher $R_{\lambda}$ results showing an inertial range for the 
magnetic helicity in the higher $k$ direct cascade region can be found in \cite{SM}.    
\begin{figure}[ht]     
   \includegraphics[width=\columnwidth]{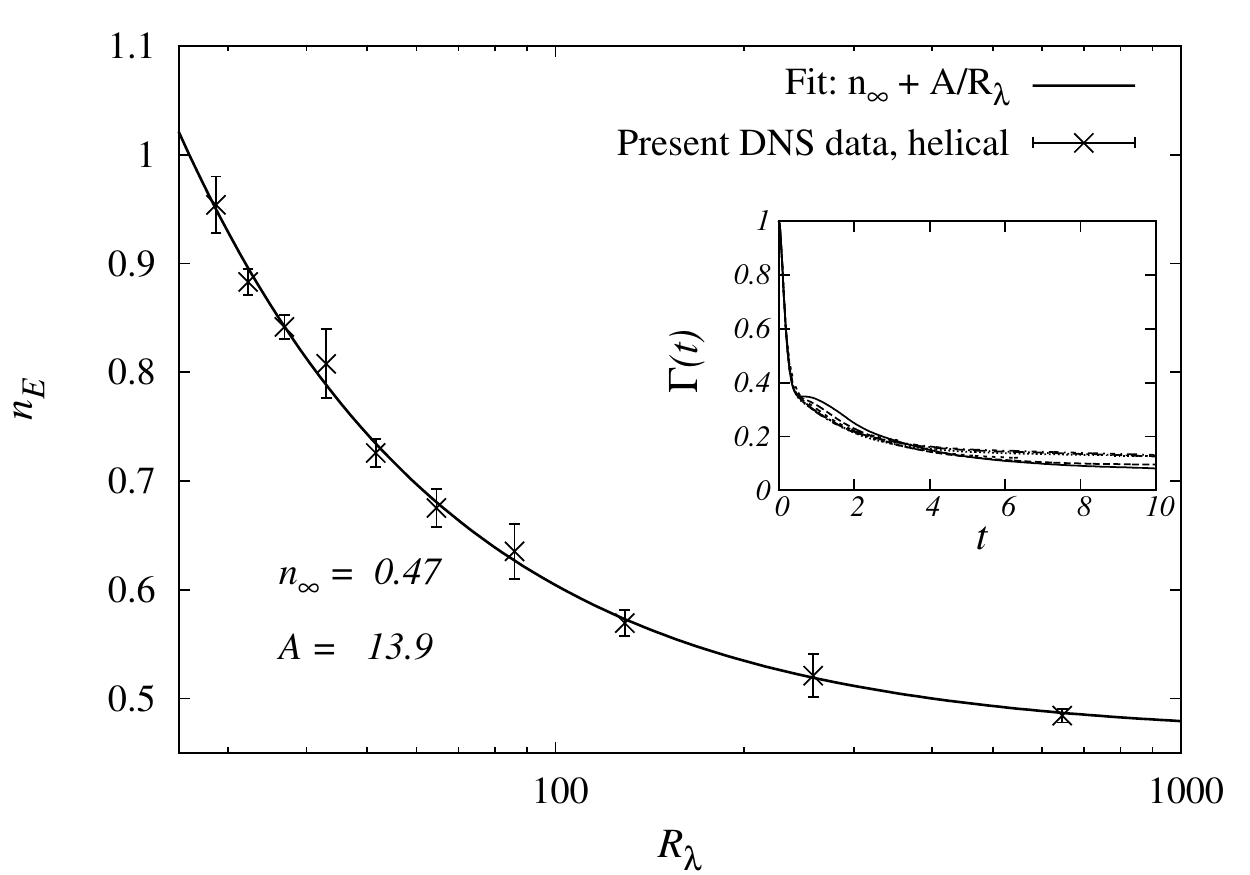} 
\caption{
Reynolds number dependence of decay exponents of $E_{mag}(t)$ for maximally 
helical initial magnetic fields.
The inset shows the decay of $\Gamma(t)$ for runs H3-H11.
 }
\label{fig:exponents}
\end{figure}

RST can also be studied through 
$E_{mag}(t)$ and $E_{kin}(t)$.
Since RST sends magnetic energy from small length scales back to large length 
scales, where dissipation is smaller, $E_{mag}(t)$ should
decay slower than $E_{kin}(t)$.  There is agreement in the MHD
literature that $E_{kin} \sim t^{-1}$ \cite{Kahniashvili12, Biskamp99, 
Christensson01, Campanelli04, Biskamp00}. 
For $E_{mag} \sim t^{-n_E}$ in 
helical MHD turbulence decay there 
are conflicting results on the 
decay exponent $n_E$, with two asymptotic decay laws proposed.
One model assumes equipartition of $E_{kin}(t)$ and $E_{mag}(t)$
during turbulence decay \cite{Biskamp99}, leading to the asymptotic
decay law $E_{mag}(t) \sim t^{-2/3}$. The second model
proposes $E_{mag} \sim E_{tot}(t) \sim t^{-1/2}$ and has been derived
in \cite{Biskamp99, Biskamp00} as an asymptotic decay law for the
\emph{total} energy with respect to late times in the decay when the
decreasing ratio $\Gamma = E_{kin}/E_{mag}$ is small, thus not assuming 
equipartition.  
Both decay laws have been observed to a good approximation for runs at 
specific Reynolds numbers 
\cite{Biskamp99, Biskamp00, Kahniashvili12, Malapaka13}. One case 
\cite{Christensson05} studied a range of low Reynolds 
numbers and attempted an extrapolation which supported the second model. 

Ensemble averaging permits a straightforward means to compute the statistical
error on the measured quantity, here $n_E$, whereas with a single realization
the only error one can obtain is the error on the fit.
Furthermore, for high resolution simulations one usually assumes that the 
ensemble average can be replaced with the volume average of one realization. 
Since RST generates long-range correlations, different regions in phase space 
will eventually become statistically correlated and the volume average will 
not reflect this. 
We observed that 
the energy spectra and the derived decay curves showed litte deviations between 
realizations for $t<7t_0$, while around $t\geqslant 7t_0$ the deviations became
significant. As an example, for run H2 $n_E$ varied from 0.81-0.96 between realizations 
if measured for $t>7t_0$. Further details can be found in \cite{SM}.

We measured $n_E$ at for a $R_{\lambda}(0)$-range of $28.69-645.47$ using ensembles
of typically 10 runs on up to $1032^3$ grid points (see Table \ref{tbl:simulations}), 
with our results in Fig.~\ref{fig:exponents}.
The largest simulation H11 was run up to $t=27t_0$, 
while the lower resolved runs reached $t=64t_0$.
As shown in the Figure, we find $n_E$ has a $1/{R_{\lambda}}$ dependence.
Extrapolating from this data to the infinite
Reynolds number limit results in an asymptotic decay law 
$E_{mag}(t) \sim t^{-n_{E,\infty}}$ with
$n_{E,\infty} = 0.47\pm 0.03$.
These results show that decay of magnetic energy in a helical system 
is slower than kinetic energy, thus supporting the presence of RST.
Moreover, our asymptote is consistent with the second model
mentioned above \cite{Biskamp99, Biskamp00} and is unambiguously 
not consistent with the first model.
As can be seen in the inset of Fig.~\ref{fig:exponents}, the ratio
$\Gamma(t)=E_{kin}(t)/E_{mag}(t)$ decreases over time.
Our results also go further than \cite{Biskamp99, Biskamp00}
as they yield finite Reynolds number results and a $1/R_{\lambda}$ 
dependence of $n_E$.
A Reynolds number dependence of $n_E$ consistent with $n_{E,\infty} = 1/2$ 
had been found before \cite{Christensson05}, albeit at much lower resolution 
using Reynolds numbers defined with respect to a lengthscale associated 
with the helicity. 
We also found that the evolution of the integral scale
approaches $L_{mag}(t) \sim t^{1/2}$ (not shown). This is consistent with the
approximate conservation of magnetic helicity at large magnetic
Reynolds number, since $H_{mag}(t)\sim E_{mag}(t)L_{mag}(t)$ \cite{Biskamp93}.

For the nonhelical case we have done 
a small analysis in response to \cite{Brandenburg14} for ensembles of 10 runs 
on up to $512^3$ grid points, resulting in exponents consistent with
$E_{mag}(t) \sim t^{-1}$, in agreement with
\cite{Kahniashvili12, Tevzadze12,Christensson01} and the theoretical analysis
by Campanelli \cite{Campanelli04}.
Since the decay exponents of $E_{kin}(t)$ and $E_{mag}(t)$ coincide for
a nonhelical magnetic system, if one field shows RST so should the other,
provided RST is large enough to influence the time evolution of the system.
Brandenburg \etal~\cite{Brandenburg14} recently reported RST
of magnetic and kinetic energies from  a single realization run of 
an initially nonhelical magnetic system on $2304^3$ grid points. 
Our ensemble of runs shows similar behavior for the magnetic energy. 

To further investigate RSTs we made an \emph{ad hoc} modification
of the momentum equation \eqref{eq:momentum} by omitting the Lorentz force
$(\nabla \times \vec{b}) \times \vec{b}$, 
which decouples the velocity field from the magnetic field.
This approach clearly does not lead to a faithful representation of MHD, 
since the decoupled fluid-magnetic field system ceases to be energetically closed.
Its purpose is to serve as a diagnostic tool to unravel the complicated
nonlinear set of equations to allow understanding the mathematical
properties of the induction equation \eqref{eq:induction} as a linear partial differential
equation. In particular, one can test if RST is among those mathematical properties.

The logic behind this modification can be viewed in another way.
One can solve the full MHD equations and store the 
$\vec{u}(\vec{x},t)$ solution.
One can imagine now doing this slightly differently by solving
just the induction equation with the same initial conditions 
using the above stored $\vec{u}(\vec{x},t)$ function. In both cases one obtains 
exactly the same solution for $\vec{b}(\vec{x},t)$. However in the second case it was
through the solution of a linear partial differential equation with 
variable coefficients, here given by $\vec{u}(\vec{x},t)$. The step
made here is to provide an alternative viewpoint, that is the entire
problem of RST can be analyzed through solving a linear partial
differential equation with variable coefficients. 
This is still a hard problem, but now linear and thus tractable.  One way forward
is to dismiss the MHD equations and study the induction equation in 
isolation with different external $\vec{u}$-fields,
as a means to probe this equation for features that produce
RST. That is what we did here with perhaps the most obvious initial example
of an independently evolving turbulent $\vec{u}$-field, and as will be shown below, 
have found even in that case, the induction equation leads to RST.

\begin{figure}[ht]     
   \includegraphics[width=\columnwidth]{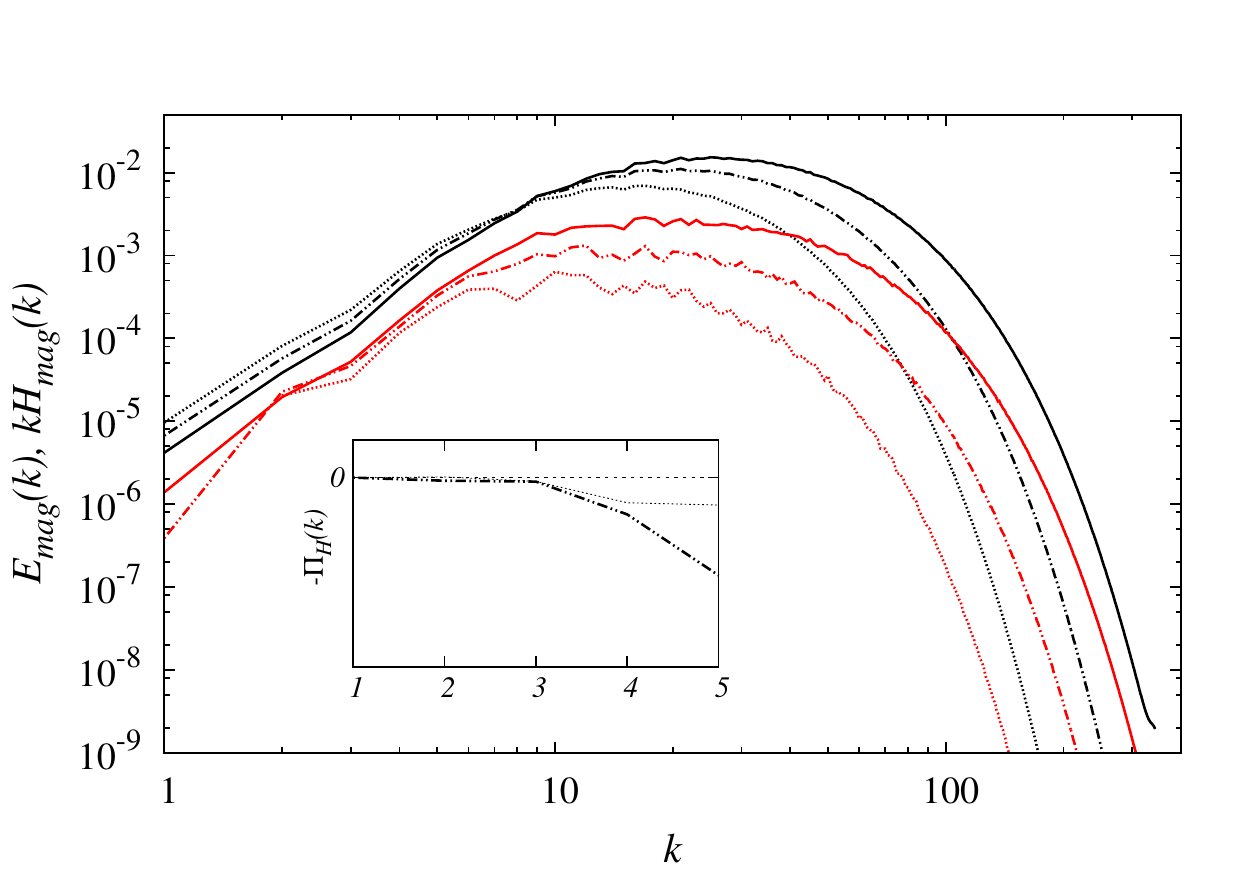} 
\caption{
(Color online) Magnetic energy spectra showing reverse spectral transfer for the decoupled system 
(run Hd5). Note the absence of RST for 
the magnetic helicity. The black (upper) lines refer to 
$E_{mag}(k,t)$, red (lower) lines to $kH_{mag}(k,t)$, solid lines 
are earlier in time than dotted lines. The inset shows the flux of magnetic 
helicity. 
}
\label{fig:ic_decoupled}
\end{figure}
We have conducted simulations for this decoupled system 
on up to $1032^3$ grid points
with our results shown in Fig.~\ref{fig:ic_decoupled} for an ensemble
of 10 runs. As seen in this Figure, there is a RST of
magnetic energy, which is interesting since it emerges from
a linear equation. To diminish the possibility of a finite-size
effect, we set the peak of the initial spectra relatively high, \emph{e.g.}~in 
Fig.~\ref{fig:ic_decoupled} $k_0=23$. 
Moreover, 
we have done several tests \cite{SM} to verify this linear RST, 
such as reproducing the same effect in slightly compressible MHD using the 
{\sc Pencil Code} \cite{pencil}. 

Interestingly,  we do not find
RST of magnetic helicity in these simulations.
The reverse transfer of $E_{mag}(k,t)$ is usually thought to be driven by the 
reverse transfer of $H_{mag}(k,t)$ by virtue of the realizability condition 
$|H_{mag}(k,t)| \leqslant 2E_{mag}(k,t)/k$ \cite{Biskamp93}.
Our results show that RST of magnetic energy is possible
without RST of magnetic helicity, despite the magnetic 
field being initially maximally helical.
Although some realizations showed RST of $H_{mag}(k,t)$ at $k=1$ and there 
appears to be a hint of RST at $k=2$, the ensemble average strongly supports 
the absence of RST of $H_{mag}(k,t)$. The data point at $k=2$ lies within the 
error of the ensemble averaged data at earlier time. 
The flux of magnetic helicity is shown in the inset of 
Fig.~\ref{fig:ic_decoupled} to be 
negative at low $k$, thus indicating the absence of RST, as opposed to the coupled 
case shown in Fig.~\ref{fig:ic}, where is it positive at low $k$. 

As the velocity field is not influenced by the magnetic field in 
this decoupled system and is initially non-helical for the results
in Fig.~\ref{fig:ic_decoupled}, 
effects of kinetic helicity could not influence the evolution 
of the magnetic field.  However having decoupled these two equations,
it permits us to cleanly test the influence that kinetic helicity can have
on the magnetic system.
\begin{figure}[ht]     
   \includegraphics[width=\columnwidth]{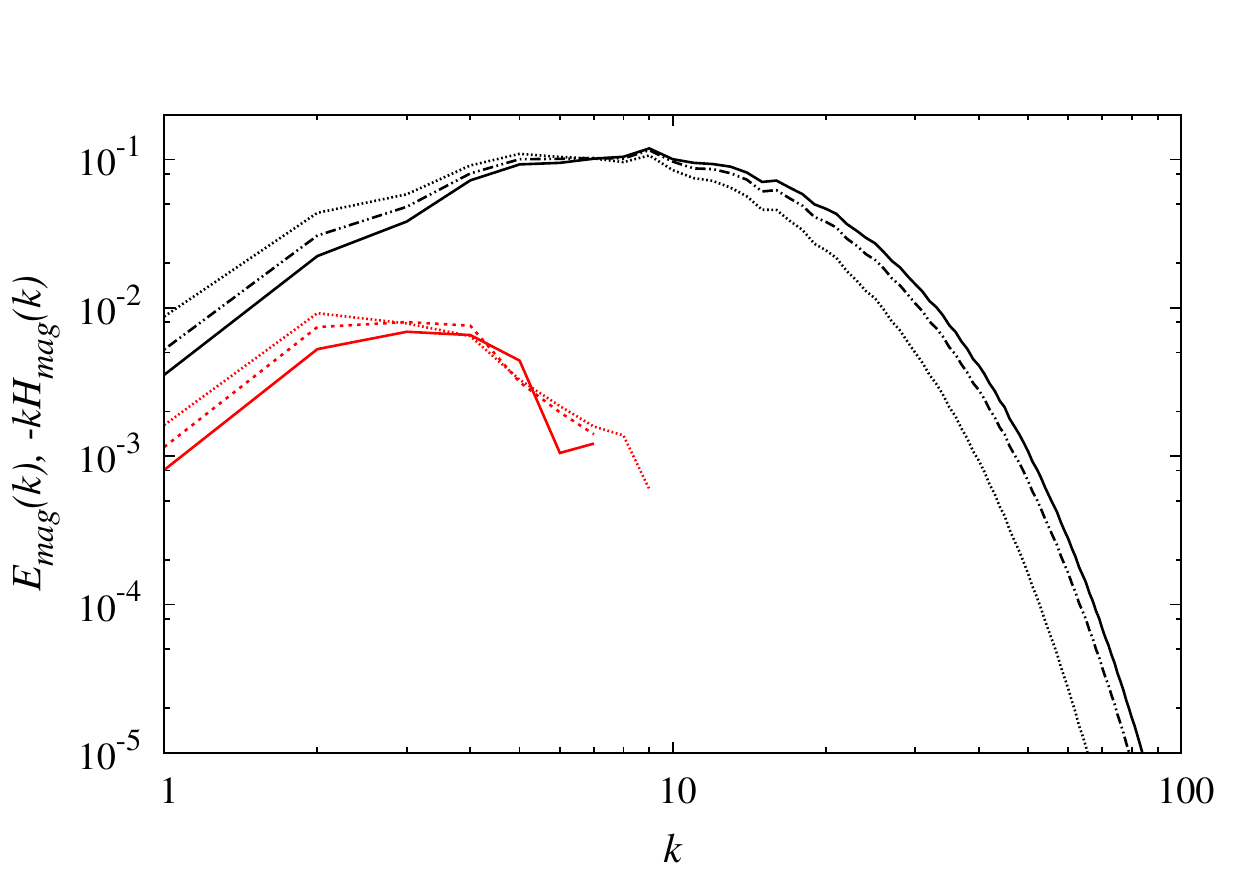} 
\caption{
(Color online) Magnetic energy and helicity spectra for the decoupled case Hd4 showing RST for an initially helical
velocity field. The black (upper) lines refer to $E_{mag}(k,t)$, 
red (lower) lines to $-kH_{mag}(k,t)$, solid lines are earlier in 
time than dotted lines. Since $H_{mag}(k,t)$ is positive at larger $k$, it does not show on logarithmic scales. 
}
\label{fig:ic_uhel}
\end{figure}
In Fig.~\ref{fig:ic_uhel} our initial $\vec{u}$- and $\vec{b}-$ fields were 
set to be maximally helical in the same direction and now we observe RST of both
magnetic energy and helicity.
In particular, this simulation was started
with $H_{mag}(k) > 0$ for all $k > 0$, and we found the
magnetic helicity to increase at large wavenumbers while it decreased at 
low wavenumbers, eventually changing sign. 
The now negative magnetic helicity 
is subsequently transferred to lower $k$. This suggests that RST of magnetic helicity 
relies on the presence of kinetic helicity, which hints at a connection
between large-scale dynamo action and RST of $H_{mag}(k,t)$. 

 \begin{table}[!Ht]
 \centering
 \begin{tabular}{lcccccc}
  Run id & $N^3$ & $R_{\lambda}(0)$ & $10^3\eta$ & $k_0$ & \#  & $t_{max}$ \\
  \hline
  \hline
  H1-8,10 & $128^3$-$528^3$ & 28.69-258.19 & 9-1 & 5-15 & 10 & 50 \\
  H9 & $1024^3$ & 74.84  & $0.75$  & 23 & 10 & 6 \\
  H11 & $1032^3$ & 645.47 & $0.4$  & 5 & 5 & 22 \\
  \hline
  NH1-6 & $128^3$-$512^3$ & 28.69-172.13 & 9-1.5 & 5 & 10 & 10-50 \\
  \hline
  Hd1-4 & $256^3$-$528^3$ & 43.03-57.38  & 6-4.5  & 5 & 10 & 5 \\
  Hd5 & $1032^3$ & 28.06 & $2$  & 23 & 10 & 2 \\
  \hline
  NHd1-3 & $256^3$-$512^3$ &43.03-57.38  & 6-4.5  & 5 & 10 & 5 \\
  \end{tabular}
 \caption{Specifications of simulations. H and NH refer to initially helical 
 and non-helical magnetic fields, respectively.
 The additional letter d refers to the decoupled system, $\eta$ denotes the
 magnetic resistivity, $k_0$ the peak wavenumber of the initial energy spectra
 and \# the ensemble size.
 }
 \label{tbl:simulations}
 \end{table}

We also examined in our decoupled equations the case of
an initially nonhelical magnetic field.
Two cases were investigated here,
one with $E_{mag}(k,0)=E_{kin}(k,0)\sim k^4$ at low $k$ and one with 
$E_{kin}(k,0)\sim k^2$ while $E_{mag}(k,0)\sim k^4$ at low $k$ as in 
\cite{Brandenburg14}. We found RST in both cases, more pronounced 
in the second than in the first case.
These results further support the findings in \cite{Brandenburg14,Zrake14} on 
nonhelical RST, now also for our linear RST.

A plausible explanation of these observations would be that RST of magnetic 
energy has two components, a dominant (non-linear) one due to the reverse
transfer of magnetic helicity and a residual 
(linear) one which is slightly augmented by the presence of magnetic helicity.  
Our numerical results show that coupling between helical modes has an impact on RST, 
especially if the coupling includes helical $\vec{u}$-modes. 
For initially helical $\vec{u}$-fields we also saw 
that coupling between positively helical $\vec{b}$- and $\vec{u}$-modes led to 
a positively helical magnetic field becoming negatively helical.   
Decomposing both fields into helical modes to study the mode 
couplings might lead to further insight. 

In order to understand the physical nature of RST, an analytic
study of the induction equation as a linear partial differential
equation in this decoupled system might lead to further insight and can
serve to get a step further towards the full nonlinear problem.
The induction equation can be further studied using
classical techniques such as Green's functions and integral transforms.
One could further disect it by retaining one of the transfer terms only. 
If the advective term $(\vec{u} \cdot \nabla)\vec{b}$ is retained, 
we obtain an advection-diffusion equation, which has been extensively studied in the literature.
The nature of the linear RST would be different depending on which of the
transfer terms produces it, and this could inspire models to be put forward 
that highlight physical processes responsible for the full nonlinear RST, 
which is analytically intractable.
In the case of the kinetic source term
$(\vec{b} \cdot \nabla)\vec{u}$, RST would be a
transfer of \emph{kinetic} to magnetic energy,
while in the case of the advective term the transfer would be of magnetic
energy only.

In summary, this paper presented the first ensemble-averaged measurements of 
reverse spectral transfer of magnetic energy and helicity, which show that these forms 
of transfer are statistical properties of the MHD equations. Our analysis 
showed that at early times single realization measurements are sufficient 
to replicate the properties of the ensemble average, while at later times 
in the decay the ensemble average becomes a must, as we observed larger 
deviations between realizations. Turbulence decay is influenced by reverse 
spectral transfer; in the helical case we 
observed a Reynolds number dependence of the decay exponent 
$n_E=n_{\infty}+A/R_{\lambda}$ with $n_{\infty} \simeq 1/2$ and 
$A=13.9 \pm 0.8$.
The reverse transfers of magnetic energy and magnetic helicity were further 
investigated in a simplified system, which decoupled the evolution of the 
velocity field from the magnetic field.  
The magnetic helicity shows no reverse transfer in this system, and
a new aspect of reverse transfer of magnetic energy was found, which is 
linear in nature and thus amenable to further mathematical analysis.

We thank David McComb for helpful advice on turbulence theory and
numerical simulations.
This work has made use of the resources provided by HECToR  
and ARCHER \cite{archer},
made available through ECDF
\cite{ecdf}. AB acknowledges funding from 
STFC, and ML is supported by 
EPSRC.

\bibliography{magn,wdm}

\begin{thebibliography}{38}
\expandafter\ifx\csname natexlab\endcsname\relax\def\natexlab#1{#1}\fi
\expandafter\ifx\csname bibnamefont\endcsname\relax
  \def\bibnamefont#1{#1}\fi
\expandafter\ifx\csname bibfnamefont\endcsname\relax
  \def\bibfnamefont#1{#1}\fi
\expandafter\ifx\csname citenamefont\endcsname\relax
  \def\citenamefont#1{#1}\fi
\expandafter\ifx\csname url\endcsname\relax
  \def\url#1{\texttt{#1}}\fi
\expandafter\ifx\csname urlprefix\endcsname\relax\def\urlprefix{URL }\fi
\providecommand{\bibinfo}[2]{#2}
\providecommand{\eprint}[2][]{\url{#2}}

\bibitem[{\citenamefont{Beresnyak}(2012)}]{Beresnyak12a}
\bibinfo{author}{\bibfnamefont{A.}~\bibnamefont{Beresnyak}},
  \bibinfo{journal}{Phys. Rev. Lett.} \textbf{\bibinfo{volume}{108}},
  \bibinfo{pages}{035002} (\bibinfo{year}{2012}).
\bibinfo{author}{\bibfnamefont{J.}~\bibnamefont{Cho}},
  \bibinfo{author}{\bibfnamefont{A.}~\bibnamefont{Lazarian}}, \bibnamefont{and}
  \bibinfo{author}{\bibfnamefont{E.~T.} \bibnamefont{Vishniac}}, in
  \emph{\bibinfo{booktitle}{Turbulence and Magnetic Fields in Astrophysics}}
  (\bibinfo{publisher}{Springer}, \bibinfo{address}{Berlin Heidelberg},
  \bibinfo{year}{2003}), pp. \bibinfo{pages}{56--98}.
\bibinfo{author}{\bibfnamefont{A.~A.} \bibnamefont{Schekochihin}}
  \bibnamefont{and} \bibinfo{author}{\bibfnamefont{S.~C.}
  \bibnamefont{Cowley}}, in
  \emph{\bibinfo{booktitle}{{Magnetohydrodynamics-Historical Evolution and
  Trends}}} (\bibinfo{publisher}{{Springer}}, \bibinfo{address}{{Berlin}},
  \bibinfo{year}{2007}), pp. \bibinfo{pages}{85--115}.

\bibitem[{\citenamefont{Brandenburg et~al.}(1996)\citenamefont{Brandenburg,
  Enqvist, and Olesen}}]{Brandenburg96}
\bibinfo{author}{\bibfnamefont{A.}~\bibnamefont{Brandenburg}},
  \bibinfo{author}{\bibfnamefont{K.}~\bibnamefont{Enqvist}}, \bibnamefont{and}
  \bibinfo{author}{\bibfnamefont{P.}~\bibnamefont{Olesen}},
  \bibinfo{journal}{Phys. Rev. D} \textbf{\bibinfo{volume}{54}},
  \bibinfo{pages}{1291} (\bibinfo{year}{1996}).

\bibitem[{\citenamefont{Son}(1999)}]{Son99}
\bibinfo{author}{\bibfnamefont{D.~T.} \bibnamefont{Son}},
  \bibinfo{journal}{Phys. Rev. D} \textbf{\bibinfo{volume}{59}},
  \bibinfo{pages}{063008} (\bibinfo{year}{1999}).

\bibitem[{\citenamefont{Tevzadze et~al.}(2012)\citenamefont{Tevzadze,
  Kisslinger, Brandenburg, and Kahniashvili}}]{Tevzadze12}
\bibinfo{author}{\bibfnamefont{A.~G.} \bibnamefont{Tevzadze}},
  \bibinfo{author}{\bibfnamefont{L.}~\bibnamefont{Kisslinger}},
  \bibinfo{author}{\bibfnamefont{A.}~\bibnamefont{Brandenburg}},
  \bibnamefont{and}
  \bibinfo{author}{\bibfnamefont{T.}~\bibnamefont{Kahniashvili}},
  \bibinfo{journal}{Astrophys J.} \textbf{\bibinfo{volume}{759}},
  \bibinfo{pages}{54} (\bibinfo{year}{2012}).

\bibitem[{\citenamefont{Kahniashvili et~al.}(2013)\citenamefont{Kahniashvili,
  Tevzadze, Brandenburg, and Neronov}}]{Kahniashvili12}
\bibinfo{author}{\bibfnamefont{T.}~\bibnamefont{Kahniashvili}},
  \bibinfo{author}{\bibfnamefont{A.~G.} \bibnamefont{Tevzadze}},
  \bibinfo{author}{\bibfnamefont{A.}~\bibnamefont{Brandenburg}},
  \bibnamefont{and} \bibinfo{author}{\bibfnamefont{A.}~\bibnamefont{Neronov}},
  \bibinfo{journal}{Phys. Rev. D} \textbf{\bibinfo{volume}{87}},
  \bibinfo{pages}{083007} (\bibinfo{year}{2013}).

\bibitem[{\citenamefont{Tobias et~al.}(2013)\citenamefont{Tobias, Cattaneo, and
  Boldyrev}}]{Tobias13}
\bibinfo{author}{\bibfnamefont{S.~M.} \bibnamefont{Tobias}},
  \bibinfo{author}{\bibfnamefont{F.}~\bibnamefont{Cattaneo}}, \bibnamefont{and}
  \bibinfo{author}{\bibfnamefont{S.}~\bibnamefont{Boldyrev}}, in
  \emph{\bibinfo{booktitle}{{Ten Chapters in Turbulence}}}
  (\bibinfo{publisher}{{Cambridge University Press}}, \bibinfo{year}{2013}),
  pp. \bibinfo{pages}{351--405}.
\bibinfo{author}{\bibfnamefont{J.}~\bibnamefont{Cho}} \bibnamefont{and}
  \bibinfo{author}{\bibfnamefont{E.~T.} \bibnamefont{Vishniac}},
  \bibinfo{journal}{Astrophys J.} \textbf{\bibinfo{volume}{538}},
  \bibinfo{pages}{217} (\bibinfo{year}{2000}).
\bibinfo{author}{\bibfnamefont{E.~T.} \bibnamefont{Vishniac}} \bibnamefont{and}
  \bibinfo{author}{\bibfnamefont{J.}~\bibnamefont{Cho}},
  \bibinfo{journal}{Astrophys J.} \textbf{\bibinfo{volume}{550}},
  \bibinfo{pages}{752} (\bibinfo{year}{2001}).

\bibitem[{\citenamefont{Iroshnikov}(1963)}]{Iroshnikov63}
\bibinfo{author}{\bibfnamefont{R.~S.} \bibnamefont{Iroshnikov}},
  \bibinfo{journal}{Sov. Astron.} \textbf{\bibinfo{volume}{7}},
  \bibinfo{pages}{566} (\bibinfo{year}{1963}).
\bibinfo{author}{\bibfnamefont{R.~H.} \bibnamefont{Kraichnan}},
  \bibinfo{journal}{Phys. Fluids} \textbf{\bibinfo{volume}{8}},
  \bibinfo{pages}{1365} (\bibinfo{year}{1965}).
\bibinfo{author}{\bibfnamefont{P.}~\bibnamefont{Goldreich}} \bibnamefont{and}
  \bibinfo{author}{\bibfnamefont{S.}~\bibnamefont{Sridhar}},
  \bibinfo{journal}{Astrophys. J.} \textbf{\bibinfo{volume}{438}},
  \bibinfo{pages}{763} (\bibinfo{year}{1995}).

\bibitem[{\citenamefont{Mininni and Pouquet}(2013)}]{Mininni13}
\bibinfo{author}{\bibfnamefont{P.~D.} \bibnamefont{Mininni}} \bibnamefont{and}
  \bibinfo{author}{\bibfnamefont{A.}~\bibnamefont{Pouquet}},
  \bibinfo{journal}{Phys. Rev. E} \textbf{\bibinfo{volume}{87}},
  \bibinfo{pages}{033002} (\bibinfo{year}{2013}).

\bibitem[{\citenamefont{Biskamp}(1993)}]{Biskamp93}
\bibinfo{author}{\bibfnamefont{D.}~\bibnamefont{Biskamp}},
  \emph{\bibinfo{title}{{Nonlinear Magnetohydrodynamics.}}}
  (\bibinfo{publisher}{Cambridge University Press}, \bibinfo{year}{1993}),
  \bibinfo{edition}{1st} ed.

\bibitem[{\citenamefont{M\"uller et~al.}(2012)\citenamefont{M\"uller, Malapaka,
  and Busse}}]{Mueller12}
\bibinfo{author}{\bibfnamefont{W.~C.} \bibnamefont{M\"uller}},
  \bibinfo{author}{\bibfnamefont{S.~K.} \bibnamefont{Malapaka}},
  \bibnamefont{and} \bibinfo{author}{\bibfnamefont{A.}~\bibnamefont{Busse}},
  \bibinfo{journal}{Phys. Rev. E} \textbf{\bibinfo{volume}{85}},
  \bibinfo{pages}{015302} (\bibinfo{year}{2012}).

\bibitem[{\citenamefont{Biskamp and M\"uller}(1999)}]{Biskamp99}
\bibinfo{author}{\bibfnamefont{D.}~\bibnamefont{Biskamp}} \bibnamefont{and}
  \bibinfo{author}{\bibfnamefont{W.-C.} \bibnamefont{M\"uller}},
  \bibinfo{journal}{Phys. Rev. Lett.} \textbf{\bibinfo{volume}{83}},
  \bibinfo{pages}{2195} (\bibinfo{year}{1999}).

\bibitem[{\citenamefont{M\"uller and Biskamp}(2000)}]{Mueller00}
\bibinfo{author}{\bibfnamefont{W.-C.} \bibnamefont{M\"uller}} \bibnamefont{and}
  \bibinfo{author}{\bibfnamefont{D.}~\bibnamefont{Biskamp}},
  \bibinfo{journal}{Phys. Rev. Lett.} \textbf{\bibinfo{volume}{84}},
  \bibinfo{pages}{03475} (\bibinfo{year}{2000}).

\bibitem[{\citenamefont{Pouquet et~al.}(1976)\citenamefont{Pouquet, Frisch, and
  L{\'e}orat}}]{Pouquet76}
\bibinfo{author}{\bibfnamefont{A.}~\bibnamefont{Pouquet}},
  \bibinfo{author}{\bibfnamefont{U.}~\bibnamefont{Frisch}}, \bibnamefont{and}
  \bibinfo{author}{\bibfnamefont{J.}~\bibnamefont{L{\'e}orat}},
  \bibinfo{journal}{J. Fluid Mech.} \textbf{\bibinfo{volume}{77}},
  \bibinfo{pages}{321} (\bibinfo{year}{1976}).

\bibitem[{\citenamefont{Yoffe}(2012)}]{Yoffe12}
\bibinfo{author}{\bibfnamefont{S.~R.} \bibnamefont{Yoffe}}, Ph.D. thesis,
  \bibinfo{school}{University of Edinburgh} (\bibinfo{year}{2012}),
  \bibinfo{note}{\url{http://arxiv.org/pdf/1306.3408v1.pdf}}.

\bibitem[{\citenamefont{Mininni et~al.}(2006)\citenamefont{Mininni, Pouquet,
  and Montgomery}}]{Mininni06}
\bibinfo{author}{\bibfnamefont{P.~D.} \bibnamefont{Mininni}},
  \bibinfo{author}{\bibfnamefont{A.~G.} \bibnamefont{Pouquet}},
  \bibnamefont{and} \bibinfo{author}{\bibfnamefont{D.~C.}
  \bibnamefont{Montgomery}}, \bibinfo{journal}{Phys. Rev. Lett.}
  \textbf{\bibinfo{volume}{97}}, \bibinfo{pages}{244503}
  (\bibinfo{year}{2006}).

\bibitem[{\citenamefont{Morales et~al.}(2014)\citenamefont{Morales, Leroy, Bos,
  and Schneider}}]{Morales12}
\bibinfo{author}{\bibfnamefont{J.~A.} \bibnamefont{Morales}},
  \bibinfo{author}{\bibfnamefont{M.}~\bibnamefont{Leroy}},
  \bibinfo{author}{\bibfnamefont{W.~J.~T.} \bibnamefont{Bos}},
  \bibnamefont{and}
  \bibinfo{author}{\bibfnamefont{K.}~\bibnamefont{Schneider}},
  \bibinfo{journal}{J. Comp. Phys.} \textbf{\bibinfo{volume}{274}},
  \bibinfo{pages}{64} (\bibinfo{year}{2014}).

\bibitem[{SM()}]{SM}
\bibinfo{howpublished}{Supplemental Material}.

\bibitem[{\citenamefont{Sagaut and Cambon}(2008)}]{Sagaut08}
\bibinfo{author}{\bibfnamefont{P.}~\bibnamefont{Sagaut}} \bibnamefont{and}
  \bibinfo{author}{\bibfnamefont{C.}~\bibnamefont{Cambon}},
  \emph{\bibinfo{title}{{Homogeneous {T}urbulence {D}ynamics}}}
  (\bibinfo{publisher}{Cambridge University Press, Cambridge, p.~95},
  \bibinfo{year}{2008}).

\bibitem[{\citenamefont{McComb}(2014)}]{McComb14a}
\bibinfo{author}{\bibfnamefont{W.~D.} \bibnamefont{McComb}},
  \emph{\bibinfo{title}{{Homogeneous, {I}sotropic {T}urbulence:
  {P}henomenology, {R}enormalization and {S}tatistical {C}losures}}}
  (\bibinfo{publisher}{Oxford University Press, p.~82}, \bibinfo{year}{2014}).

\bibitem[{\citenamefont{Christensson et~al.}(2001)\citenamefont{Christensson,
  Hindmarsh, and Brandenburg}}]{Christensson01}
\bibinfo{author}{\bibfnamefont{M.}~\bibnamefont{Christensson}},
  \bibinfo{author}{\bibfnamefont{M.}~\bibnamefont{Hindmarsh}},
  \bibnamefont{and}
  \bibinfo{author}{\bibfnamefont{A.}~\bibnamefont{Brandenburg}},
  \bibinfo{journal}{Phys. Rev. E} \textbf{\bibinfo{volume}{64}},
  \bibinfo{pages}{056405} (\bibinfo{year}{2001}).

\bibitem[{\citenamefont{Campanelli}(2004)}]{Campanelli04}
\bibinfo{author}{\bibfnamefont{L.}~\bibnamefont{Campanelli}},
  \bibinfo{journal}{Phys. Rev. D} \textbf{\bibinfo{volume}{70}},
  \bibinfo{pages}{083009} (\bibinfo{year}{2004}).

\bibitem[{\citenamefont{Biskamp and M\"uller}(2000)}]{Biskamp00}
\bibinfo{author}{\bibfnamefont{D.}~\bibnamefont{Biskamp}} \bibnamefont{and}
  \bibinfo{author}{\bibfnamefont{W.-C.} \bibnamefont{M\"uller}},
  \bibinfo{journal}{Phys. Plasma} \textbf{\bibinfo{volume}{7}},
  \bibinfo{pages}{4889} (\bibinfo{year}{2000}).

\bibitem[{\citenamefont{Malapaka and M\"uller}(2013)}]{Malapaka13}
\bibinfo{author}{\bibfnamefont{S.~K.} \bibnamefont{Malapaka}} \bibnamefont{and}
  \bibinfo{author}{\bibfnamefont{W.-C.} \bibnamefont{M\"uller}},
  \bibinfo{journal}{Astrophys J.} \textbf{\bibinfo{volume}{778}}
  (\bibinfo{year}{2013}).

\bibitem[{\citenamefont{Christensson et~al.}(2005)\citenamefont{Christensson,
  Hindmarsh, and Brandenburg}}]{Christensson05}
\bibinfo{author}{\bibfnamefont{M.}~\bibnamefont{Christensson}},
  \bibinfo{author}{\bibfnamefont{M.}~\bibnamefont{Hindmarsh}},
  \bibnamefont{and}
  \bibinfo{author}{\bibfnamefont{A.}~\bibnamefont{Brandenburg}},
  \bibinfo{journal}{Astron. Nachr.} \textbf{\bibinfo{volume}{326}},
  \bibinfo{pages}{393} (\bibinfo{year}{2005}).

\bibitem[{\citenamefont{Brandenburg et~al.}(2014)\citenamefont{Brandenburg,
  Kahniashvili, and Tevzadze}}]{Brandenburg14}
\bibinfo{author}{\bibfnamefont{A.}~\bibnamefont{Brandenburg}},
  \bibinfo{author}{\bibfnamefont{T.}~\bibnamefont{Kahniashvili}},
  \bibnamefont{and} \bibinfo{author}{\bibfnamefont{A.~G.}
  \bibnamefont{Tevzadze}},
  \bibinfo{howpublished}{\url{http://arxiv.org/abs/1404.2238}}
  (\bibinfo{year}{2014}).

\bibitem[{pen()}]{pencil}
\bibinfo{howpublished}{\url{http://pencil-code.googlecode.com/}}.

\bibitem[{\citenamefont{Zrake}(2014)}]{Zrake14}
\bibinfo{author}{\bibfnamefont{J.}~\bibnamefont{Zrake}},
  \bibinfo{howpublished}{\url{http://arxiv.org/pdf/1407.5626v1.pdf}}
  (\bibinfo{year}{2014}).

\bibitem[{arc()}]{archer}
\bibinfo{howpublished}{\url{http://www.hector.ac.uk/}}.
\bibinfo{howpublished}{\url{http://www.archer.ac.uk/}}.

\bibitem[{ecd()}]{ecdf}
\bibinfo{howpublished}{\url{http://www.ecdf.ed.ac.uk/}}.

\end{thebibliography}


\begin{thebibliography}{6}%
\makeatletter
\providecommand \@ifxundefined [1]{%
 \@ifx{#1\undefined}
}%
\providecommand \@ifnum [1]{%
 \ifnum #1\expandafter \@firstoftwo
 \else \expandafter \@secondoftwo
 \fi
}%
\providecommand \@ifx [1]{%
 \ifx #1\expandafter \@firstoftwo
 \else \expandafter \@secondoftwo
 \fi
}%
\providecommand \natexlab [1]{#1}%
\providecommand \enquote  [1]{``#1''}%
\providecommand \bibnamefont  [1]{#1}%
\providecommand \bibfnamefont [1]{#1}%
\providecommand \citenamefont [1]{#1}%
\providecommand \href@noop [0]{\@secondoftwo}%
\providecommand \href [0]{\begingroup \@sanitize@url \@href}%
\providecommand \@href[1]{\@@startlink{#1}\@@href}%
\providecommand \@@href[1]{\endgroup#1\@@endlink}%
\providecommand \@sanitize@url [0]{\catcode `\\12\catcode `\$12\catcode
  `\&12\catcode `\#12\catcode `\^12\catcode `\_12\catcode `\%12\relax}%
\providecommand \@@startlink[1]{}%
\providecommand \@@endlink[0]{}%
\providecommand \url  [0]{\begingroup\@sanitize@url \@url }%
\providecommand \@url [1]{\endgroup\@href {#1}{\urlprefix }}%
\providecommand \urlprefix  [0]{URL }%
\providecommand \Eprint [0]{\href }%
\@ifxundefined \urlstyle {%
  \providecommand \doi  [0]{\begingroup \@sanitize@url \@doi}%
  \providecommand \@doi [1]{\endgroup \@@startlink {\doibase
  #1}doi:\discretionary {}{}{}#1\@@endlink }%
}{%
  \providecommand \doi  [0]{doi:\discretionary{}{}{}\begingroup
  \urlstyle{rm}\Url }%
}%
\providecommand \doibase [0]{http://dx.doi.org/}%
\providecommand \Doi [0]{\begingroup \@sanitize@url \@Doi }%
\providecommand \@Doi  [1]{\endgroup\@@startlink{\doibase#1}\@@Doi}%
\providecommand \@@Doi [1]{#1\@@endlink}%
\providecommand \selectlanguage [0]{\@gobble}%
\providecommand \bibinfo  [0]{\@secondoftwo}%
\providecommand \bibfield  [0]{\@secondoftwo}%
\providecommand \translation [1]{[#1]}%
\providecommand \BibitemOpen [0]{}%
\providecommand \bibitemStop [0]{}%
\providecommand \bibitemNoStop [0]{.\EOS\space}%
\providecommand \EOS [0]{\spacefactor3000\relax}%
\providecommand \BibitemShut  [1]{\csname bibitem#1\endcsname}%
\bibitem [{\citenamefont {Berera}\ and\ \citenamefont
  {Linkmann}(2014)}]{Berera14}%
  \BibitemOpen
  \bibfield  {author} {\bibinfo {author} {\bibfnamefont {A.}~\bibnamefont
  {Berera}}\ and\ \bibinfo {author} {\bibfnamefont {M.~F.}\ \bibnamefont
  {Linkmann}},\ }\href@noop {} { (\bibinfo {year} {2014})}\BibitemShut
  {NoStop}%
\bibitem [{\citenamefont {Mininni}\ \emph {et~al.}(2006)\citenamefont
  {Mininni}, \citenamefont {Pouquet},\ and\ \citenamefont
  {Montgomery}}]{Mininni06}%
  \BibitemOpen
  \bibfield  {author} {\bibinfo {author} {\bibfnamefont {P.~D.}\ \bibnamefont
  {Mininni}}, \bibinfo {author} {\bibfnamefont {A.~G.}\ \bibnamefont
  {Pouquet}}, \ and\ \bibinfo {author} {\bibfnamefont {D.~C.}\ \bibnamefont
  {Montgomery}},\ }\href@noop {} {\bibfield  {journal} {\bibinfo  {journal}
  {Phys. Rev. Lett.},\ }\textbf {\bibinfo {volume} {97}},\ \bibinfo {pages}
  {244503} (\bibinfo {year} {2006})}\BibitemShut {NoStop}%
\bibitem [{\citenamefont {Morales}\ \emph {et~al.}(2014)\citenamefont
  {Morales}, \citenamefont {Leroy}, \citenamefont {Bos},\ and\ \citenamefont
  {Schneider}}]{Morales12}%
  \BibitemOpen
  \bibfield  {author} {\bibinfo {author} {\bibfnamefont {J.~A.}\ \bibnamefont
  {Morales}}, \bibinfo {author} {\bibfnamefont {M.}~\bibnamefont {Leroy}},
  \bibinfo {author} {\bibfnamefont {W.~J.~T.}\ \bibnamefont {Bos}}, \ and\
  \bibinfo {author} {\bibfnamefont {K.}~\bibnamefont {Schneider}},\ }\href@noop
  {} {\bibfield  {journal} {\bibinfo  {journal} {J. Comp. Phys.},\ }\textbf
  {\bibinfo {volume} {274}},\ \bibinfo {pages} {64} (\bibinfo {year}
  {2014})}\BibitemShut {NoStop}%
\bibitem [{\citenamefont {M\"uller}\ \emph {et~al.}(2012)\citenamefont
  {M\"uller}, \citenamefont {Malapaka},\ and\ \citenamefont
  {Busse}}]{Mueller12}%
  \BibitemOpen
  \bibfield  {author} {\bibinfo {author} {\bibfnamefont {W.~C.}\ \bibnamefont
  {M\"uller}}, \bibinfo {author} {\bibfnamefont {S.~K.}\ \bibnamefont
  {Malapaka}}, \ and\ \bibinfo {author} {\bibfnamefont {A.}~\bibnamefont
  {Busse}},\ }\href@noop {} {\bibfield  {journal} {\bibinfo  {journal} {Phys.
  Rev. E},\ }\textbf {\bibinfo {volume} {85}},\ \bibinfo {pages} {015302}
  (\bibinfo {year} {2012})}\BibitemShut {NoStop}%
\bibitem [{Pen()}]{Pencil}%
  \BibitemOpen
  \href@noop {} {}\bibinfo {howpublished}
  {\url{http://pencil-code.googlecode.com/}}\BibitemShut {NoStop}%
\bibitem [{\citenamefont {Mininni}\ and\ \citenamefont
  {Pouquet}(2013)}]{Mininni13}%
  \BibitemOpen
  \bibfield  {author} {\bibinfo {author} {\bibfnamefont {P.~D.}\ \bibnamefont
  {Mininni}}\ and\ \bibinfo {author} {\bibfnamefont {A.}~\bibnamefont
  {Pouquet}},\ }\href@noop {} {\bibfield  {journal} {\bibinfo  {journal} {Phys.
  Rev. E},\ }\textbf {\bibinfo {volume} {87}},\ \bibinfo {pages} {033002}
  (\bibinfo {year} {2013})}\BibitemShut {NoStop}%
\end{thebibliography}%

\end{document}


\title{Supplemental Material for "Magnetic helicity and the 
evolution of decaying magnetohydrodynamic turbulence"}
\author{Arjun Berera and Moritz Linkmann}
\affiliation{SUPA, School of Physics and Astronomy, University of Edinburgh, UK}


\begin{abstract}
In this supplemental material we provide details of our investigation
of reverse spectral transfers in magnetohydrodynamic turbulence, such as 
verification of our code and further tests of our results.  
\end{abstract}
\maketitle

\section{Introduction}
In the paper \cite{Berera14} we conducted direct numerical simulations (DNS) 
of incompressible magnetohydrodynamic (MHD) turbulence without an imposed guide field.
The MHD equations for incompressible flow are
\begin{align}
\label{eq:momentum}
\partial_t \vec{u}&= - \frac{1}{\rho}\nabla P -(\vec{u}\cdot \nabla)\vec{u}
 + \frac{1}{\rho}(\nabla \times \vec{b}) \times \vec{b} + \nu \Delta \vec{u}  \ , \\
\label{eq:induction}
\partial_t \vec{b}&= (\vec{b}\cdot \nabla)\vec{u}-(\vec{u}\cdot \nabla)\vec{b} + \eta \Delta \vec{b}\ , \\
\label{eq:incompr}
&\nabla \cdot \vec{u} = 0 \ \ \mbox{and} \ \  \nabla \cdot \vec{b} = 0 \ ,  
\end{align}
where $\vec{u}$ denotes the velocity field, $\vec{b}$ the magnetic induction 
expressed in Alfv\'{e}n units, $\nu$ the kinematic viscosity, $\eta$ the 
resistivity, $P$ the pressure and $\rho=1$ the density.

We presented ensemble averaged data showing reverse spectral transfer (RST) of magnetic
helicity and measured decay exponents for initially maximally helical 
magnetic fields.
In a particular set of simulations we decoupled the momentum equation 
\eqref{eq:momentum} from the induction equation \eqref{eq:induction} and showed that the inverse
cascade of magnetic energy persists, while the reverse spectral transfer of magnetic 
helicity disappears in the decoupled system, provided the initial velocity field was nonhelical. 

Details of our simulations can be found in Table \ref{tbl:simulations}, where we specify values 
for Reynolds numbers, resolution, diffusivities, runtime $t_{max}$, ensemble size 
and wavenumber $k_0$ at which we set the peaks of the initial spectra.

\section{Verification of DNS code}
\begin{figure}[!t]     
        \begin{center}
 \subfigure[ Evolution of the maximal current density]{
   \includegraphics[width=0.48\textwidth]{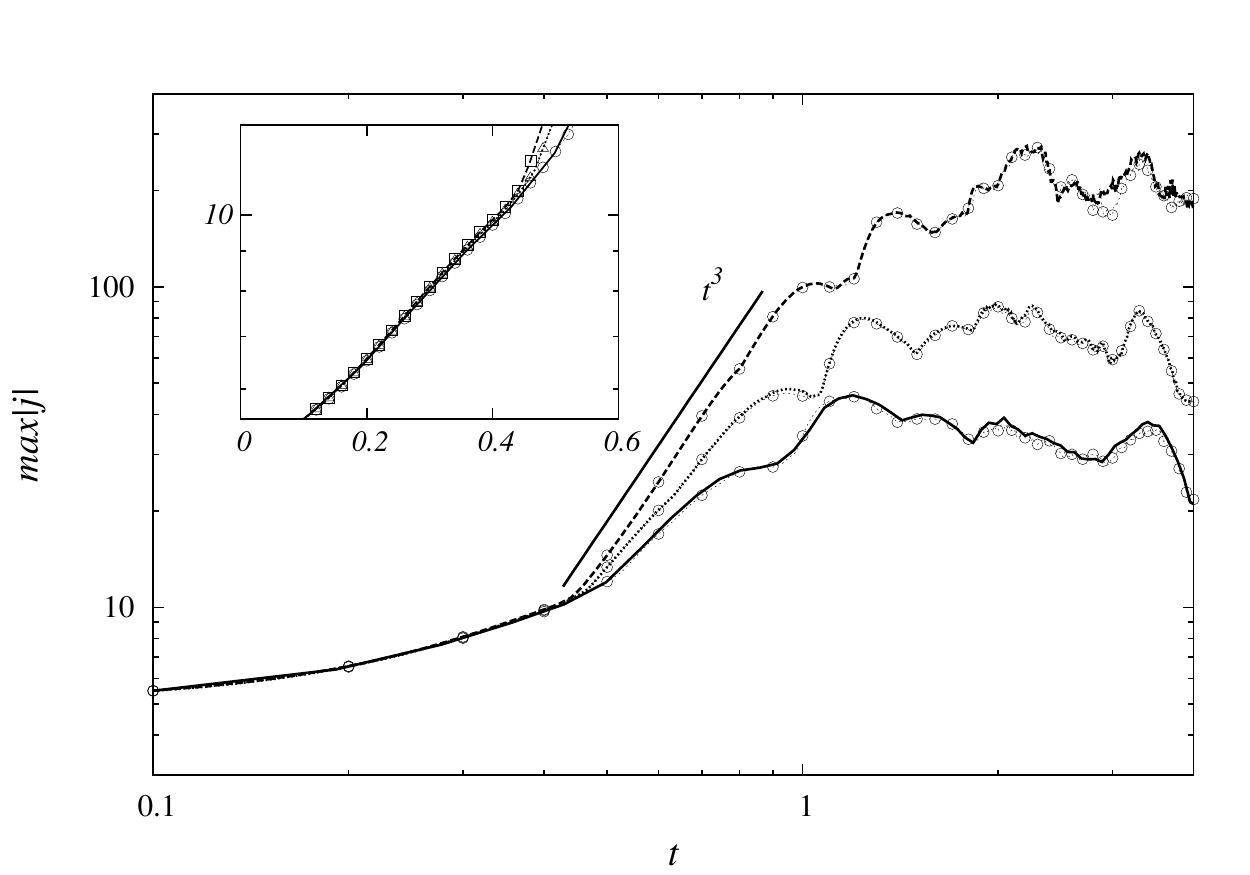}
   }
 \subfigure[ Magnetic energy]{
   \includegraphics[width=0.48\textwidth]{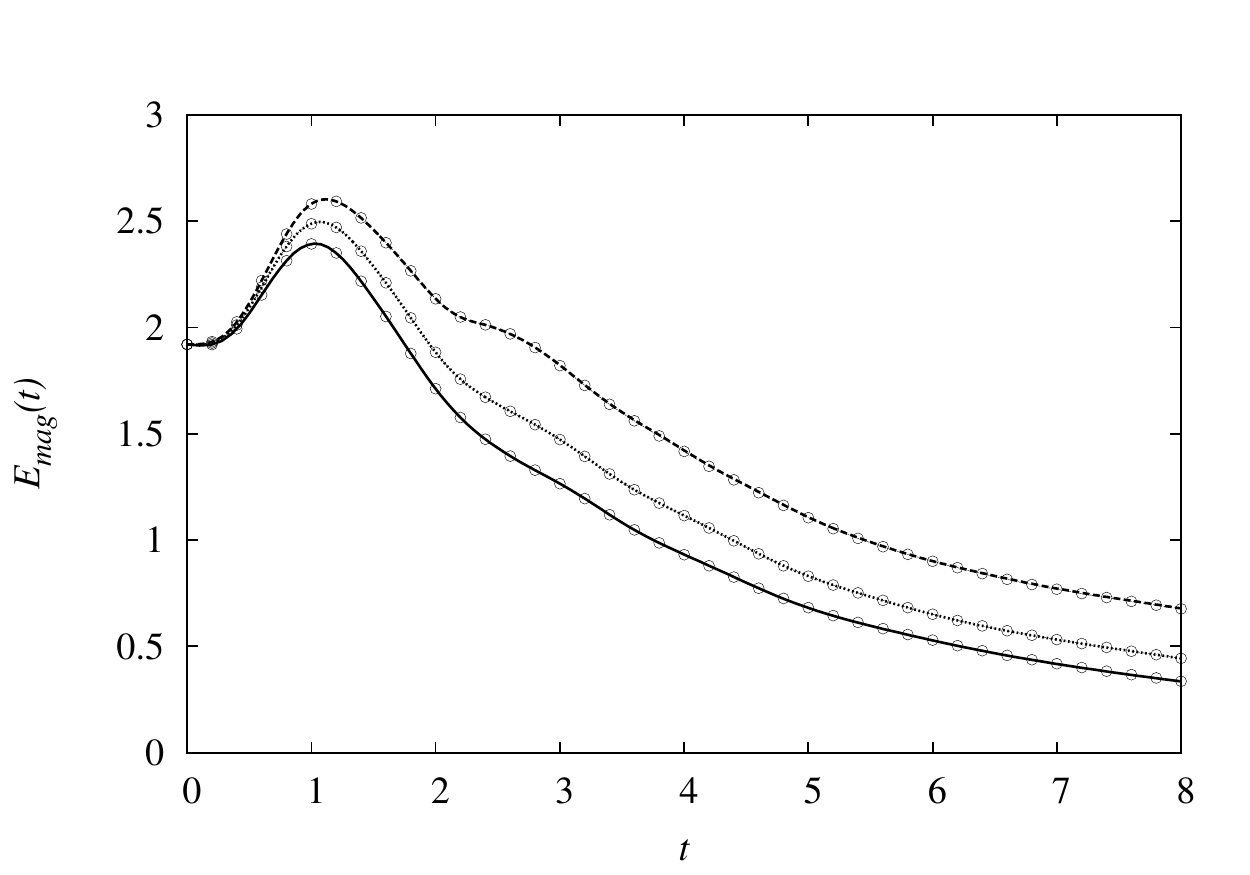}
   }
\end{center}
\caption{Comparison to Morales \etal~The symbols refer to our DNS, the lines to DNS by Morales \etal}
\label{fig:Morales_comp}
\end{figure}

The 3D Orszag-Tang vortex was used for comparison against results by Mininni, 
Pouquet and Montgomery \cite{Mininni06} and Morales, Leroy, Bos and Schneider \cite{Morales12}.
The initial conditions are
    \begin{align}
    {\vec{u}}&= (-2 \sin{y}, \ 2 \sin{x},\ 0) \nonumber \ , \\
    {\vec{b}}&= \beta (-2 \sin{2y} + \sin{z}, \ 2 \sin{x} + \sin{z}, \ \sin{x} + \sin{y}) \ ,  
    \label{eq:OT_init}
    \end{align} 
  where $\beta=0.8$ has been chosen according to \cite{Mininni06, Morales12}. 
  We calculate the total dissipation
   \begin{equation}
\varepsilon_{tot}(t)= \nu \langle \vec{\omega}^2 \rangle + \eta \langle \vec{j}^2 \rangle \ ,
   \end{equation}
and the maximum of the current density in real space $\mbox{max}|{\vec{j}}|$.
We used the same number of grid points and diffusivities as \cite{Morales12},
that is $\nu=\eta =0.01$ on $64^3$ points, $\nu=\eta =0.005$ on $128^3$ points
and $\nu=\eta =0.001$ on $256^3$ points.
Our results are in agreement with both above mentioned sources. 
The maximum of the current density shows Reynolds number independent 
exponential growth until $t=0.4$, in agreement with Morales \etal~, while 
Mininni \etal~observe exponential growth up to $t=0.6$. After this initial period we 
observe algebraic growth $\sim t^3$ in agreement with both sources. With 
increasing Reynolds number the temporal maxima of $\mbox{max}|{\bf{j}}|$ are 
achieved at later times, also in agreement with both 
sources.
Our data shows very good agreement to a dataset obtained from 
Morales \etal~\cite{Morales12}, see Fig.~\ref{fig:Morales_comp}.  
\section{Inertial range properties for the full MHD equations.} 
\begin{figure}[!t]     
        \begin{center}
   \includegraphics[width=0.7\textwidth]{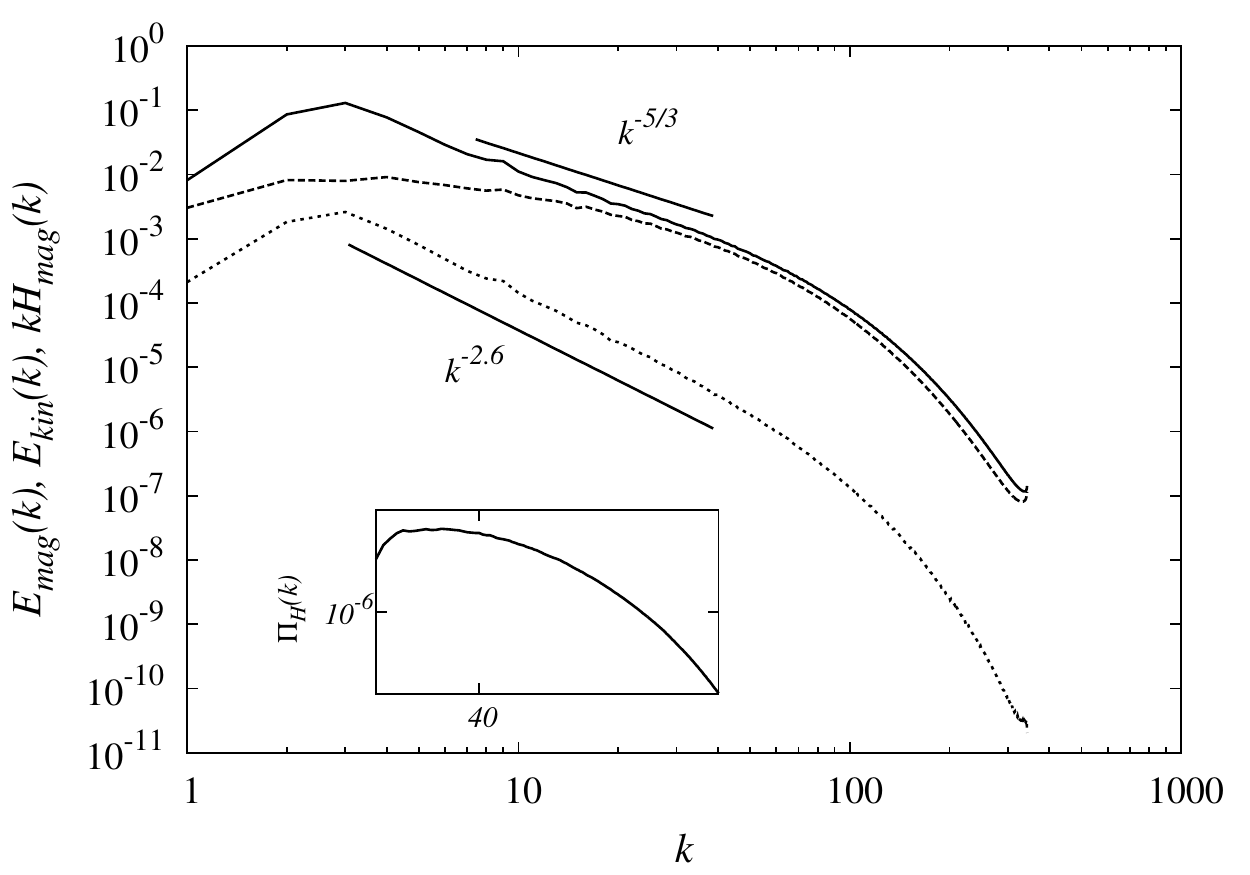}
\end{center}
\caption{
Spectra for the largest Reynolds number run (H11). The solid line shows $E_{mag}(k)$, 
the middle dashed line shows $E_{kin}(k)$ and the
bottom dotted line shows $kH_{mag}(k)$, which has been shifted for easier comparison. 
The inset shows the flux of $H_{mag}(k)$, which is constant in the
higher $k$ region, indicating an inertial range. The straight lines indicate scaling regions 
for $E_{mag}(k)\sim k^{-5/3}$ and $kH_{mag}(k)\sim k^{-2.6}$, which results in 
$H_{mag}(k)\sim k^{-3.6}$.
}
\label{fig:inertial}
\end{figure}

Figure \ref{fig:inertial} shows magnetic helicity, magnetic energy and kinetic energy spectra
shortly after the onset of power-law decay (that is, at $t=1.3t_0$) for the largest Reynolds 
number run H11. The magnetic helicity spectra have been multiplied by the wavenumber $k$ 
for dimensional reasons, and have been 
shifted downwards in the figure to facilitate visual analysis. The inset shows 
constancy of helicity flux $\Pi_H(k)$ for the wavenumber interval $21 \leqslant k \leqslant 33$, 
indicating an inertial range for the magnetic helicity in the direct cascade region. 
It can be seen in the figure that power-law scaling of $H_{mag}(k)$ extends over a larger interval, 
showing the scaling $H_{mag}(k)\sim k^{-3.6}$, which is in agreement with 
recent results on decaying 3D MHD turbulence \cite{Mueller12}.
The flux is $k$-dependent in the reverse spectral transfer 
region, as shown in the paper in the inset of Fig.~1 for the lower $R_{\lambda}$-run H9 and 
here in the inset of Fig.~\ref{fig:ic} for run H4a. 
This is also in agreement with \cite{Mueller12}, who report the same behavior.
Our simulations support Kolmogorov scaling for the magnetic field as indicated 
by the bar parallel to $E_{mag}(k)$.  

\section{Evolution of the low wavenumber form of kinetic energy spectra for the full MHD equations.}
\begin{figure}[!t]     
        \begin{center}
   \includegraphics[width=0.7\textwidth]{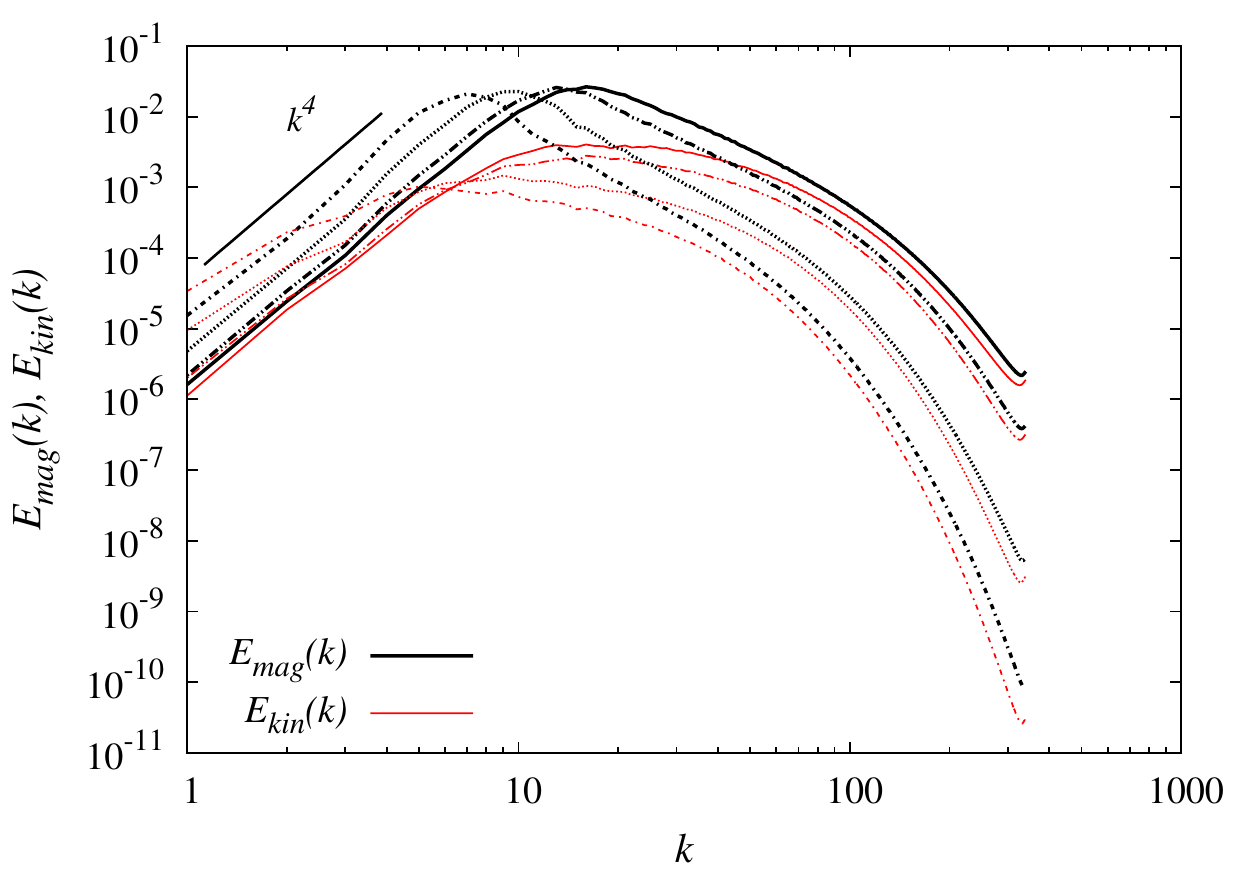}
\end{center}
\caption{
(Color online) Evolution of kinetic (thin red (grey) lines) and magnetic (thick black lines) energy spectra over time for run H9. 
Note how the slope of $E_{kin}(k)$ flattens with time, while the slope of $E_{mag}(k)$ 
remains unchanged. 
}
\label{fig:lowk}
\end{figure}

Figure \ref{fig:lowk} shows kinetic and magnetic energy spectra on 
logarithmic scales at different times during the decay for run H9, which is the run with the 
largest scale separation. Note that the slope of $E_{kin}(k)$ becomes flatter over time while 
the slope of $E_{mag}(t)$ remains the same, that is $E_{mag}(k) \sim k^4$ at small $k$ prevails, 
as indicated the diagonal line superimposed above $E_{mag}(k)$ in the 
low wavenumber region. The slope of $E_{kin}(k)$
does \emph{not} change from $k^4$ to $k^2$, instead it continues to flatten over time.

\section{Further tests of linear RST}
\begin{figure}[!t]     
        \begin{center}
   \subfigure[ Comparison to hydrodynamics]{
   \label{fig:hydro}
   \includegraphics[width=0.48\textwidth]{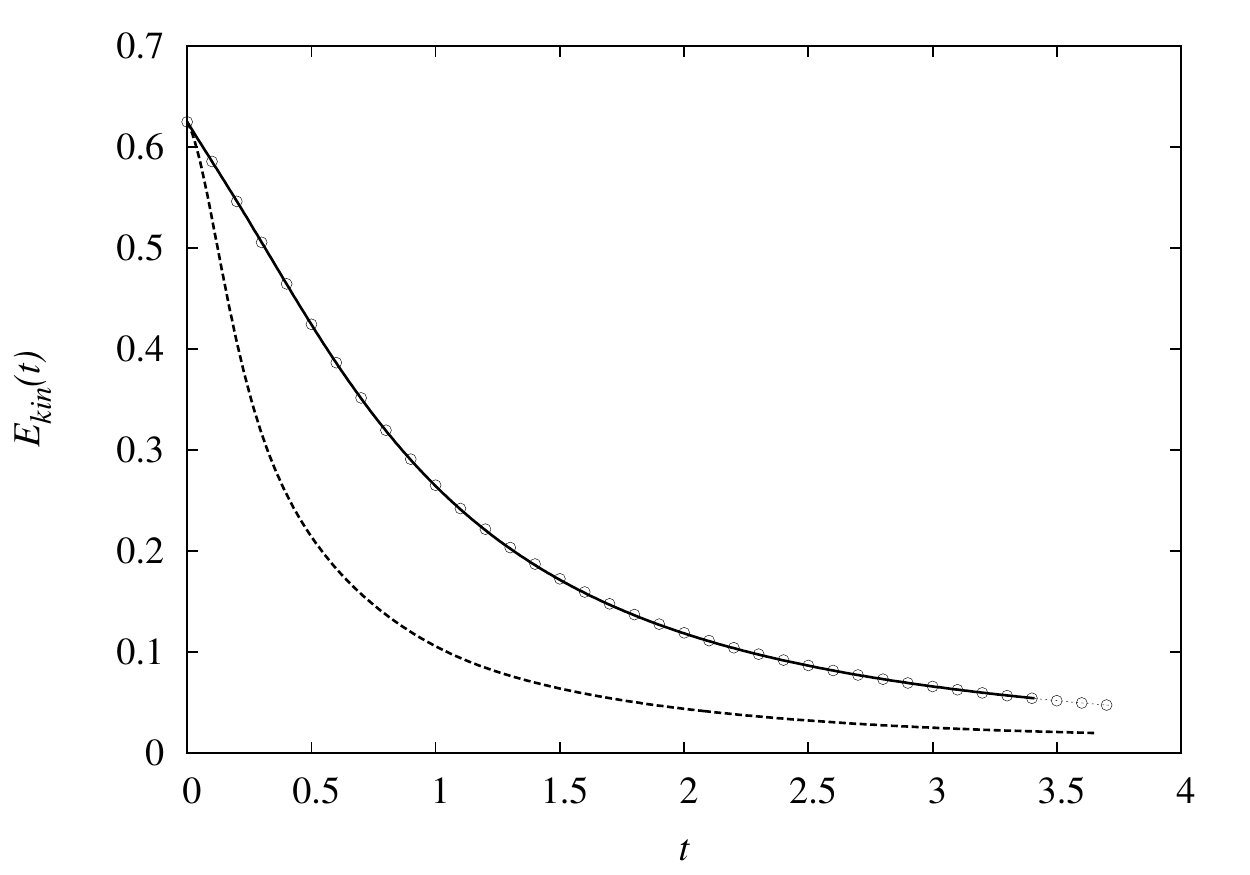}
   }
   \subfigure[ Results using the {\sc Pencil Code}]{
   \label{fig:pencil}
   \includegraphics[width=0.48\textwidth]{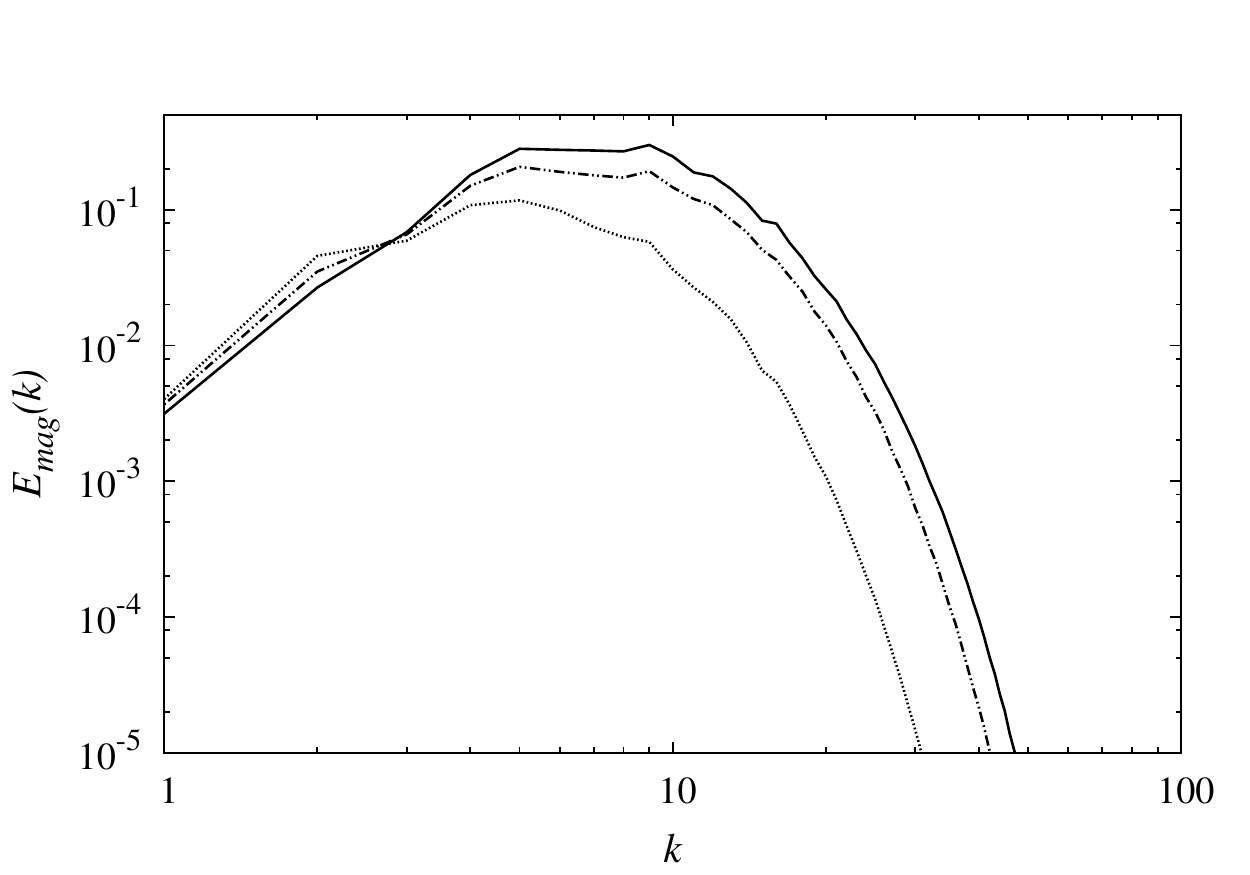}
   }
\end{center}
\caption{(a) The solid line shows $E_{kin}(t)$ 
from a hydrodynamic run, the symbols shows $E_{kin}(t)$ from a 
simulation where the momentum equation was decoupled from the 
induction equation, and the dashed line shows $E_{kin}(t)$ of 
a full MHD run. The evolution of $E_{kin}(t)$ coincides for the
hydrodynamic run and the simulation of the decoupled case, showing that
in the decoupled case the velocity field evolves independently 
of the magnetic field, as expected. 
(b) Linear RST of magnetic energy using the {\sc Pencil Code} on $128^3$ 
grid points.
 }
\end{figure}

We have conducted further tests in order to validate the linear RST found in our simulations. The velocity field should evolve as in hydrodynamic turbulence if 
the Lorentz force term is omitted from the momentum equation. We have tested 
that this is the case, see Fig.~\ref{fig:hydro}, this test clearly shows that
the velocity field evolves independently from the magnetic field.
\begin{figure}[!t]     
        \begin{center}
 \subfigure[ Coupled case]{
\label{fig:ic}
   \includegraphics[width=0.48\textwidth]{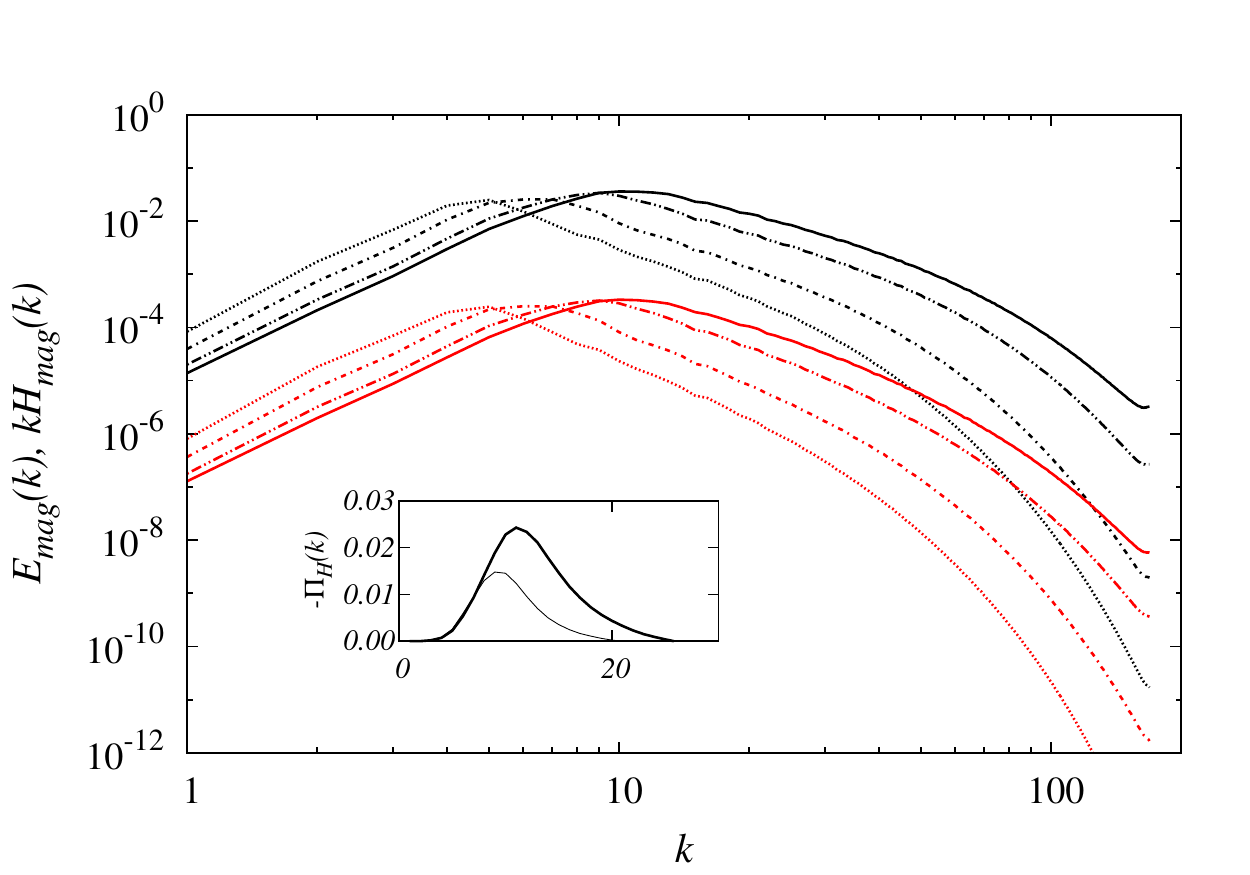}
   }
 \subfigure[ Decoupled case]{
  \label{fig:ic_decoupled}
   \includegraphics[width=0.48\textwidth]{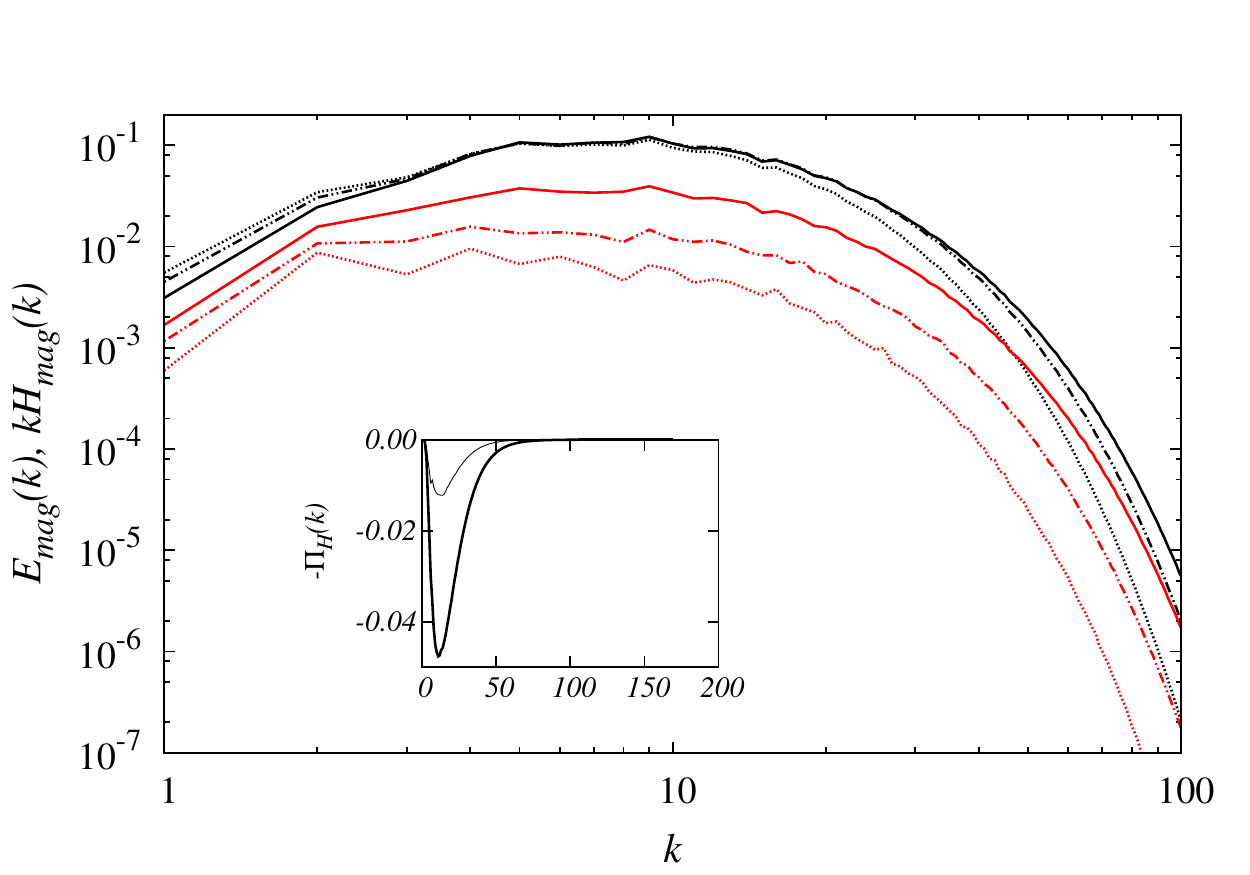}
   }
\end{center}
\caption{(a) (Color online) Magnetic energy and helicity spectra showing reverse spectral transfer for a run on $512^3$ grid points (H4a). 
The black (upper) lines refer to $E_{mag}(k)$, red (lower) lines refer to 
$H_{mag}(k)$. Solid lines indicate one initial large eddy
turnover time, dotted and dash-dotted 
lines refer to 2, 5 and 10 turnover times. The inset shows the 
flux of magnetic helicity at one and two turnover times. 
(b) Magnetic energy and helicity spectra for the decoupled case on $512^3$
grid points (Hd3). Upper (black) lines refer to $E_{mag}(k)$, 
lower (red) lines to $H_{mag}(k)$. The inset shows the flux of magnetic helicity. 
         }
\end{figure}
The linear RST of magnetic energy  was also found for slightly 
compressible MHD using the 
{\sc Pencil Code} \cite{Pencil}. 
Figure \ref{fig:pencil} shows that after removing the Lorentz force term 
in the momentum equation in the {\sc Pencil Code} we still observe RST of 
magnetic energy. 

Results for the linear RST of magnetic energy were found in simulations 
using $256^3$ up to $1032^3$ collocation 
points, but only the highest resolved results with the largest scale separation 
are included in the paper \cite{Berera14}. Figure \ref{fig:ic_decoupled} shows 
results from simulations 
using $512^3$ collocation points (Hd3), where the peaks of the initial spectra had 
been set at $k_0 = 5$. The linear RST is still visible, but concentrated at the 
very lowest wavenumbers. In contrast, the simulation results
shown in the paper are obtained by setting the peaks of the initial spectra at 
$k_0=23$, and RST can be seen up to $k=7$ in Fig. 3(a) of the paper.
Figure \ref{fig:ic_decoupled} also shows that RST is absent for the magnetic
helicity, as there is no increase of $H_{mag}(k)$ at the low wavenumbers and the flux
of magnetic helicity is negative. For comparison, the coupled case shows RST for both 
$H_{mag}(k)$ and $E_{mag}(k)$ as can be seen in Fig.~\ref{fig:ic}, where the inset shows
positive flux of kinetic helicity in the low $k$ region.

\begin{figure}[!t]     
        \begin{center}
   \subfigure[ Energy spectra]{
   \includegraphics[width=0.4\textwidth]{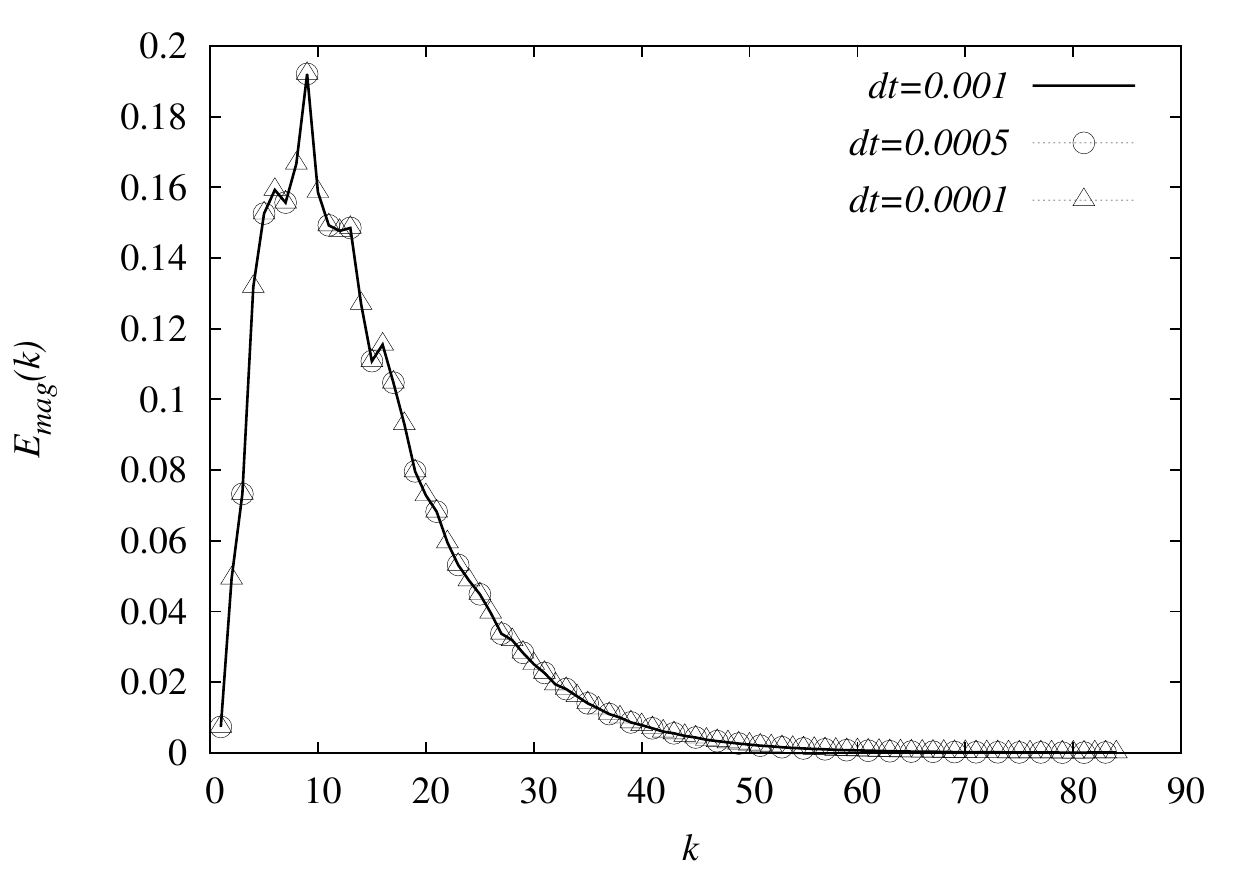}
    }
   \subfigure[ Time evolution of the magnetic energy]{
   \includegraphics[width=0.4\textwidth]{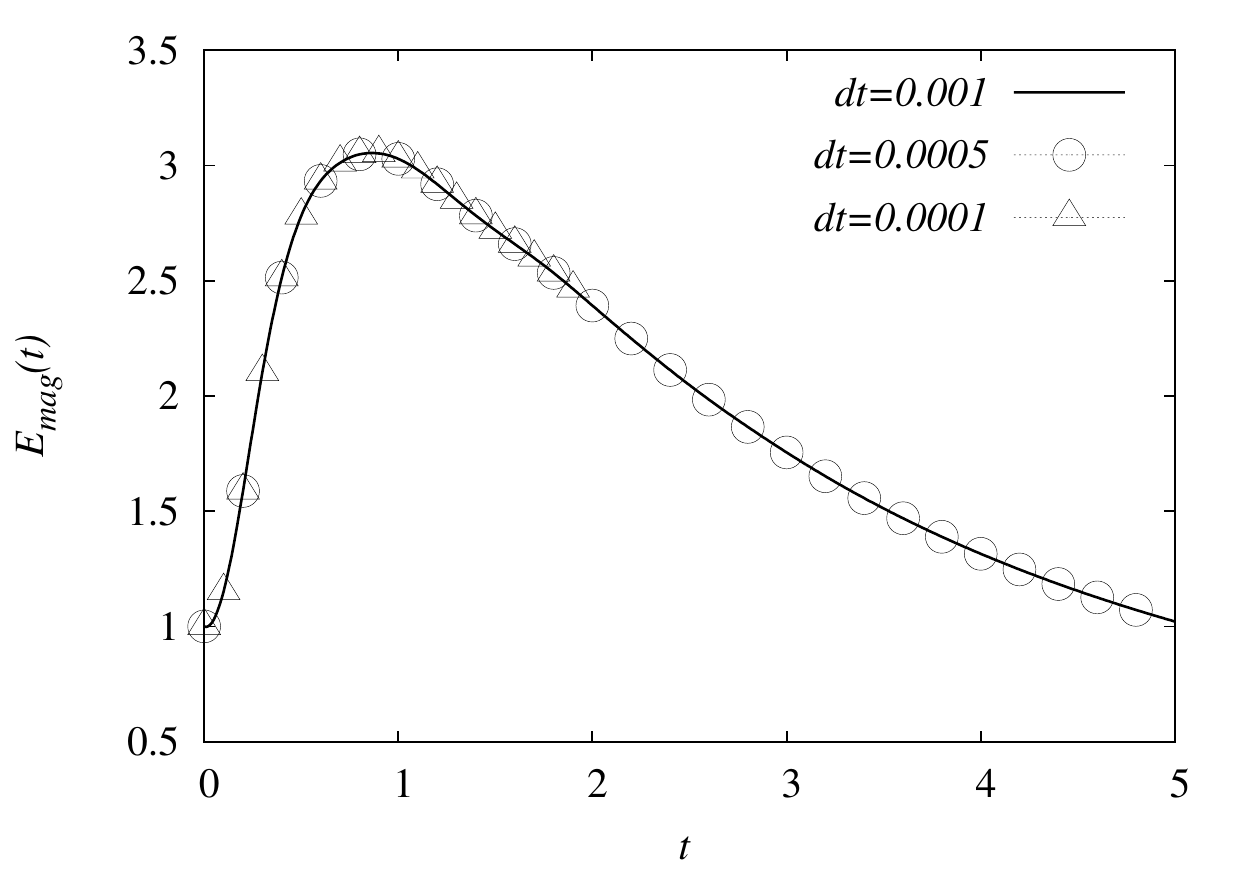}
    }
\end{center}
\caption{
(a) Results for magnetic energy spectra from simulations run at different
time-steps on $256^3$ grid points. The spectra obtained from simulations using
different time-steps coincide.
(b) The evolution for the magnetic energy is the same for simulations using 
different time-steps.
}
\label{fig:dec_convergence}
\end{figure}

Figure \ref{fig:dec_convergence} shows that our simulations have converged for 
the time-step chosen. We do not see differences in the magnetic energy spectra
 and the magnetic energy as we reduce the time-step, results are shown for 
$256^3$ grid points. 

\section{Importance of ensemble averaging}
\subsection{Decay exponents}
\begin{figure}[!t]     
        \begin{center}
   \includegraphics[width=0.8\textwidth]{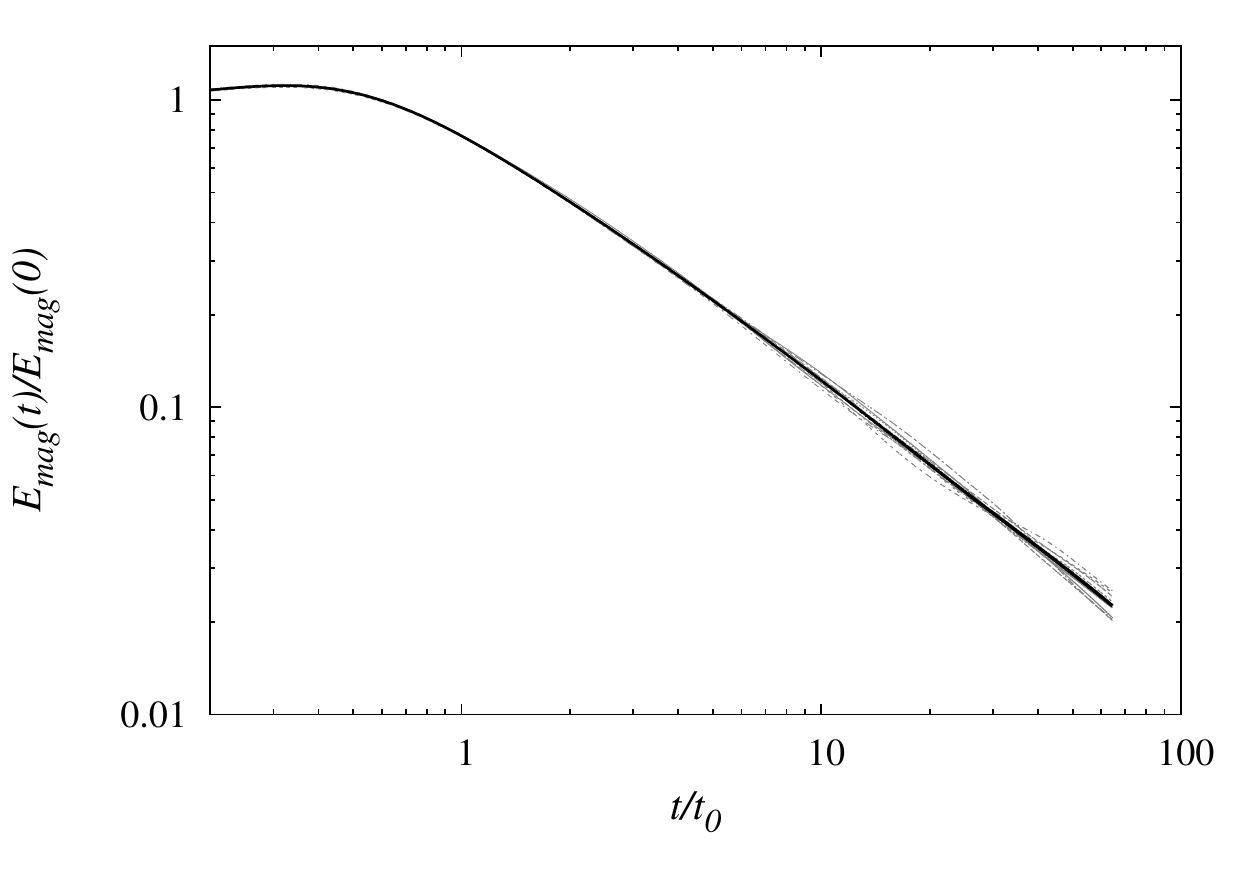}
\end{center}
\caption{
Deviation of realizations (grey lines) from the ensemble average (black line)
of run H2.
Note that there is a clearly visible spread between realizations for 
$t/t_0 \geq 7$. 
}
\label{fig:EA_exponents}
\end{figure}

Figure ~\ref{fig:EA_exponents} illustrates the importance of ensemble
averaging for the measurement of the decay exponent $n_E$, 
showing all realizations of H2 and the ensemble average.
It can clearly be seen how the decay exponents for single realizations
differ at later times, and how they deviate from the exponent measured at 
early times. By contrast, the ensemble average follows the same straight line
for all times after the onset of power-law decay.
Similarly, a recent study of 2D hydrodynamic turbulence \cite{Mininni13} 
using ensemble averaging showed a significant spread in the evolution of the 
integral scale for different realizations and a somewhat less pronounded 
spread in the total energy decay curves (see their Fig.~3). 
The decay exponent $n_E$ measured
from individual realizations of H2 can range from 0.81-0.96 for $t>7t_0$,
where $t_0$ denotes the initial large eddy turnover time.

\subsection{Magnetic helicity for the decoupled case}
\begin{figure}[H]     
        \begin{center}
   \subfigure[ Helicity spectra of all realizations of the ensemble Hd5]{
   \includegraphics[width=0.8\textwidth]{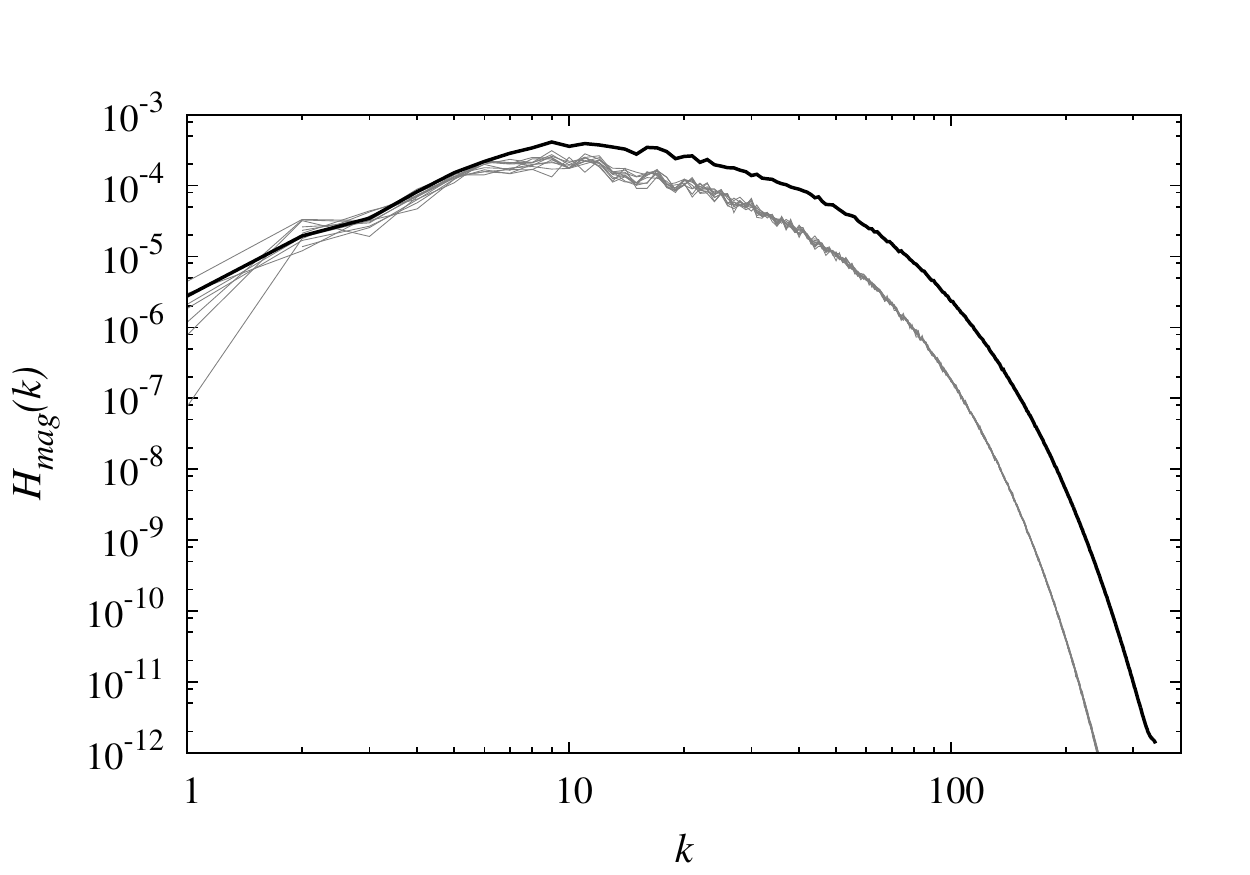}
\label{fig:EA_decIC}
    }
   \subfigure[ Ensemble averaged helicity spectra for run Hd5]{
   \includegraphics[width=0.8\textwidth]{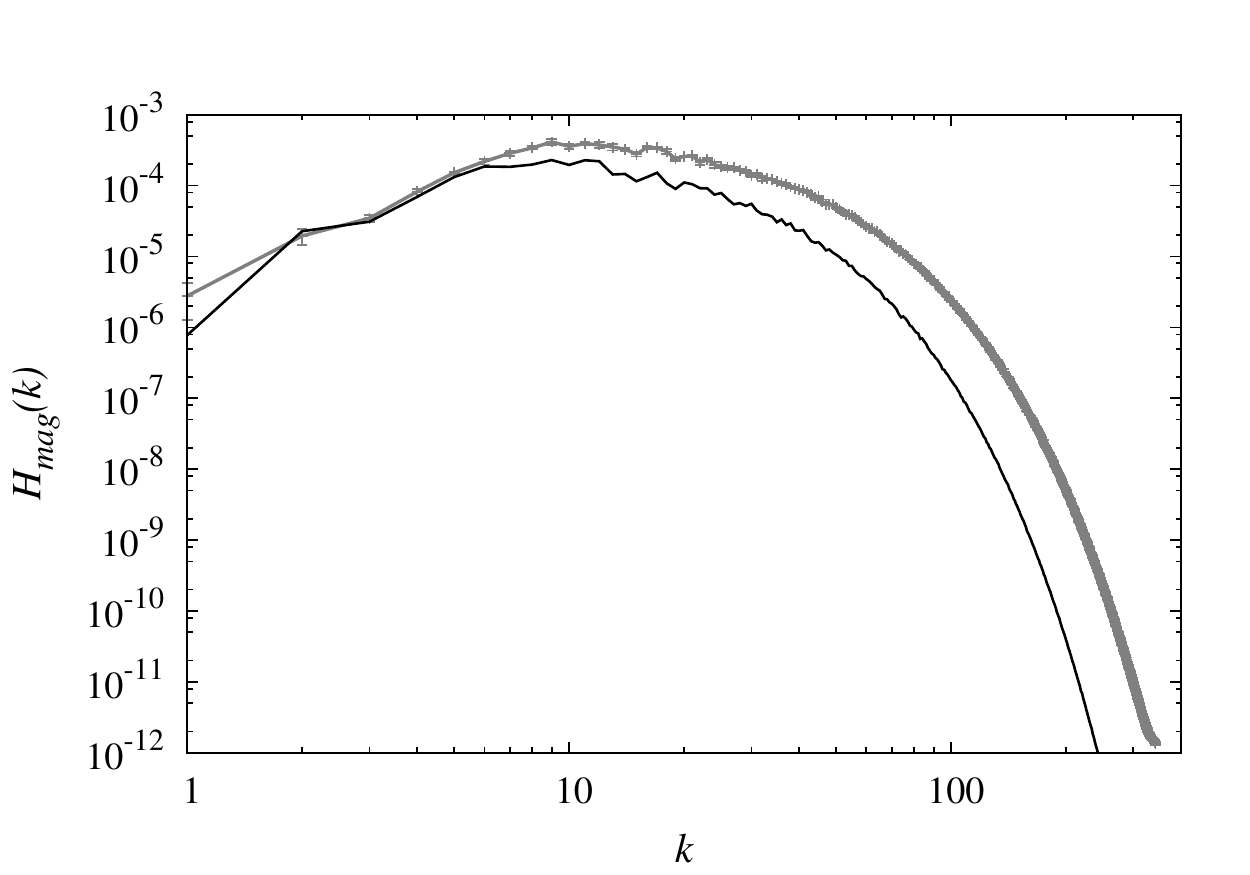}
\label{fig:EA_dec_err}
    }
\end{center}
\caption{
(a) The black line shows the ensemble average of $H_{mag}(k)$, the grey lines 
show $H_{mag}(k)$ for all realizations of run Hd5 at later time. Note the
large deviations between realizations at low $k$.
(b) The black line shows the ensemble average over the realizations in (a). 
What appears as a hint of RST at $k=2$ lies within the error of the ensemble
average at earlier time (grey line). 
}
\end{figure}

As mentioned in the paper \cite{Berera14}, some realizations of Hd5 appeared
to show RST, but the ensemble average did not confirm this. As can be seen
in Fig.~\ref{fig:EA_decIC}, the error on the ensemble averaged helicity 
spectrum is large at $k=1$. Some realizations clearly show RST, some do not and
some realizations show negative helicity at $k=1$, which is why the value
does not appear on a logarithmic scale.

Taking data from one realization only could have lead to erroneous results, 
claiming that either $H_{mag}(k)$ does show reverse spectral transfer, or that 
$H_{mag}(k)$ becomes negative at $k=1$. The ensemble averaged result shows that
neither is the case. For $k=2$ there appears to be a hint of RST, but as can
be seen in Fig.~\ref{fig:EA_dec_err} the helicity spectrum at later time lies 
within the error of the helicity spectrum at earlier time.

We would like to thank Jorge Morales for kindly providing us with a dataset of 
simulation results using the Orszag-Tang vortex as initial condition.

 \begin{table}[H]
 \centering
 \begin{tabular}{ccccccccc}
   Run id & $N^3$  & $k_{max}\eta_{mag}$ & $R_{\lambda}(0) $  &  $\eta $  & $k_{0}$ & \#  & $t_{max}/s$ \\
  \hline
  \hline
  H1 & $128^3$ & 1.30 & 28.69 & $0.009$ & 5 & 10 & 50 \\
  H2 & $256^3$ & 2.42 & 32.27  & $0.008$ &  5 & 10 & 50\\
  H3 & $256^3$ & 2.18 & 36.88 & $0.007$  & 5 & 10 & 50\\
  H4 & $256^3$ & 2.08 & 43.03 & $0.006$  & 5 & 10 & 50 \\
  H5 & $256^3$ & 1.80 & 51.64 & 0.005  &  5 & 10 & 50  \\
  H6 &  $256^3$ & 1.59 & 64.55 & 0.004 &  5 & 10 &50  \\
  H7 & $256^3$ & 1.30 & 86.06  & $0.0031$  & 5 & 10 & 50 \\
  H4a & $512^3$ & 1.38 & 43.03  & $0.002$  & 15 & 10 & 48 \\
  H8 & $512^3$ & 2.01 & 129.09  & $0.002$ & 5 & 10 & 7 \\
  H9 & $1024^3$ & 1.38 & 74.84  & $0.00075$  & 23 & 10 & 6 \\
  H10 & $528^3$ & 1.31 & 258.19  & $0.001$  & 5 & 10 & 50 \\
  H11 & $1032^3$ & 1.38 & 645.47  & $0.0004$  & 5 & 5 & 12 \\
  \hline
  NH1 & $128^3$ & 1.29 & 28.69  & $0.009$ & 5 & 10 & 50 \\
  NH2 & $256^3$ & 1.51 & 64.55 & $0.004$  & 5 & 10 & 10 \\
  NH3 & $256^3$ & 1.26 & 86.06  & $0.003$ & 5 & 10 & 12 \\
  NH4 & $512^3$ & 1.43 & 129.09  & $0.002$  & 5 & 10 & 10 \\
  NH5 & $512^3$ & 1.30 & 172.13  & $0.0015$  & 5 & 10 & 7 \\
  \hline
  Hd1 & $256^3$ & 1.26 & 43.03  & $0.006$  & 5 & 10 & 5 \\
  Hd2 & $256^3$ & 1.26 & 43.03  & $0.006$  & 5 & 10 & 5 \\
  Hd3 & $512^3$ & 1.88 & 57.38  & $0.0045$  & 5 & 10 & 5 \\
  Hd4 & $528^3$ & 1.95 & 57.38  & $0.0045$  & 5 & 10 & 5 \\
  Hd5 & $1032^3$ & 1.39 & 28.06 & $0.001$  & 23 & 10 & 2 \\
  \hline
  NHd1 & $256^3$ & 1.26 & 43.03  & $0.006$  & 5 & 10 & 5 \\
  NHd2 & $256^3$ & 1.26 & 43.03  & $0.006$  & 5 & 10 & 5 \\
  NHd3 & $512^3$ & 1.88 & 57.38  & $0.0045$  & 5 & 10 & 5 \\
  \end{tabular}
 \caption{Specifications of simulations. H refers to an initially helical
 magnetic field, NH to an initially non-helical magnetic field.
 The additional letter d refers to the decoupled system, $R_{\lambda}(0)$ 
 denotes the initial Taylor-scale Reynolds number, $\eta$ the
 magnetic resistivity, $k_0$ the peak wavenumber of the initial energy spectra,
 $k_{max}$ the largest resolved wavenumber, $\eta_{mag}$ the Kolmogrov microscale 
 associated with the magnetic field, \# the ensemble size and $t_{max}$ the 
 run time. The run NHd2 was 
 started from initial spectra following $k^2$ at low wavenumbers $k$, and  
 runs Hd2 and Hd4 were started using a maximally helical initial velocity field. 
 }
 \label{tbl:simulations}
 \end{table}

    
\bibliography{magn,wdm}